\documentclass[10pt,journal,twocolumn]{IEEEtran}

\usepackage{cite}

\ifCLASSINFOpdf
  \usepackage[pdftex]{graphicx}
\else
  \usepackage[dvips]{graphicx}
\fi

\usepackage[cmex10]{amsmath}

\usepackage{algorithmic,algorithm}
\usepackage[tight,footnotesize]{subfigure}
\usepackage{DefineNotations}
\usepackage{amsfonts}
\usepackage{amsbsy,amssymb}
\usepackage{color}

\begin{document}
%
\title{Sum Rate Maximized Resource Allocation in Multiple DF Relays Aided OFDM Transmission}
\author{Tao Wang and Luc Vandendorpe,~\IEEEmembership{Fellow, IEEE}
\thanks{
Manuscript was received on October 8, 2010, and revised on February 14, 2011.
T. Wang and L. Vandendorpe are with ICTEAM Institute,
Universit\'e Catholique de Louvain, 1348 Louvain-la-Neuve, Belgium
(email: \{tao.wang,luc.vandendorpe\}@uclouvain.be).}
\thanks{
The authors would like to thank the Walloon Region for funding the projects
MIMOCOM, the ARC SCOOP and the FP7 Network NEWCOM++.
}
}
%
%

\markboth{To appear in IEEE Journal on Selected Areas in Communications}%
{}
%



\maketitle

\begin{abstract}
In relay-aided wireless transmission systems, one of the key issues is
how to manage the energy resource at the source and each individual relay,
to optimize a certain performance metric.
This paper addresses the sum rate maximized resource allocation (RA) problem
in an orthogonal frequency division modulation (OFDM) transmission system
assisted by multiple decode-and-forward (DF) relays,
subject to the individual sum power constraints of the source and the relays.
In particular, the transmission at each subcarrier can be in either
the direct mode without any relay assisting,
or the relay-aided mode with one or several relays assisting.
We propose two RA algorithms which optimize the assignment of transmission mode and source power
for every subcarrier, as well as the assisting relays and the power allocation to them
for every {relay-aided} subcarrier.
First, it is shown that the considered RA problem has zero Lagrangian duality gap
when there is a big number of subcarriers.
In this case, a duality based algorithm that finds a globally optimum RA is developed.
Most interestingly, the sensitivity analysis in convex optimization theory
is used to derive a closed-form optimum solution to a related convex optimization problem,
for which the method based on the Karush-Kuhn-Tucker (KKT) conditions is not applicable.
Second, a coordinate-ascent based iterative algorithm, which finds a suboptimum RA but is always applicable
regardless of the duality gap of the RA problem, is developed.
The effectiveness of these algorithms has been illustrated by numerical experiments.
\end{abstract}


%
\IEEEpeerreviewmaketitle

\section{Introduction}

{Relay-aided} cooperative transmission finds plenty of promising applications
when it is difficult to install multiple antennas at the same radio equipment,
and therefore has been attracting intensive research interest in both academia and industry lately \cite{Pabst04}.
Low complexity yet efficient protocols,
such as amplify and forward (AF) as well as decode and forward (DF), have been
proposed to simplify the implementation with practical devices \cite{Laneman03,Laneman04}.
Typically, both protocols propose to carry out a {relay-aided} transmission within two time slots,
namely a broadcasting slot and a relaying slot.
In \cite{Laneman04}, the AF/DF which fixes every transmission in the {relay-aided} mode
independently of source-relay channel conditions is referred to as fixed relaying AF/DF.
In fact, improved protocols may be built for better performance.
For instance, selection relaying AF/DF,
which selects either the direct or {relay-aided} transmission mode
depending on channel conditions, has been proposed to improve spectral efficiency \cite{Laneman04}.
In particular, the direct mode, which refers to the direct source to destination transmission
without any relay assisting, is used when the source-destination channel gain
is higher than the source-relay channel gain.
Most interestingly, not fixed but selection relaying DF can achieve full diversity \cite{Laneman04}.

We consider in this paper a point to point orthogonal frequency division modulation (OFDM)
transmission system aided by multiple DF relays.
The motivation behind this is that the OFDM transmission has been widely recognized for current and future wireless systems,
thanks to its flexibility to incorporate dynamic resource allocation (RA) for performance improvement \cite{wang11}.
In such a transmission system, one of the key issues is how to
decide for every subcarrier the transmission mode and assisting relays
if the {relay-aided} mode is chosen,
and the power of the source and every individual relay,
to maximize a certain objective related to system performance.
Obviously, this RA problem is more complicated compared to those for conventional OFDM systems without relays.
Therefore, RA algorithms are solicited for {relay-aided} OFDM systems.

To date, some related research works have been reported.
To name a few, RA algorithms have been proposed in \cite{Jang10,Pham10,Dang10,Duval10}
for OFDM systems aided by AF relays,
and in \cite{Nam07,Kaneko07,Kwak07,Ng07,Cui09,Salem10} for multi-user OFDM systems aided by DF relays.
As for OFDM systems aided by DF relays,
RA algorithms have been proposed in \cite{Gui08} to minimize sum power under rate constraints,
and in \cite{Li08,Vandendorpe08-1} to maximize sum rate
subject to power constraints, when a single relay exists.
However, at a subcarrier in the direct mode the source is idle during the relaying slot,
which wastes spectrum resource.
To address this issue, rate-optimized RA algorithms, which allow for
the source to destination transmission during the two slots at the subcarrier in the direct mode,
have been proposed in \cite{Vandendorpe08-2,Vandendorpe09-1,Vandendorpe09-2,Vandendorpe09-3}.

\begin{figure}[!t]
  \centering
  \subfigure[]{
     \includegraphics[width=1.3in, height = 1.5in]{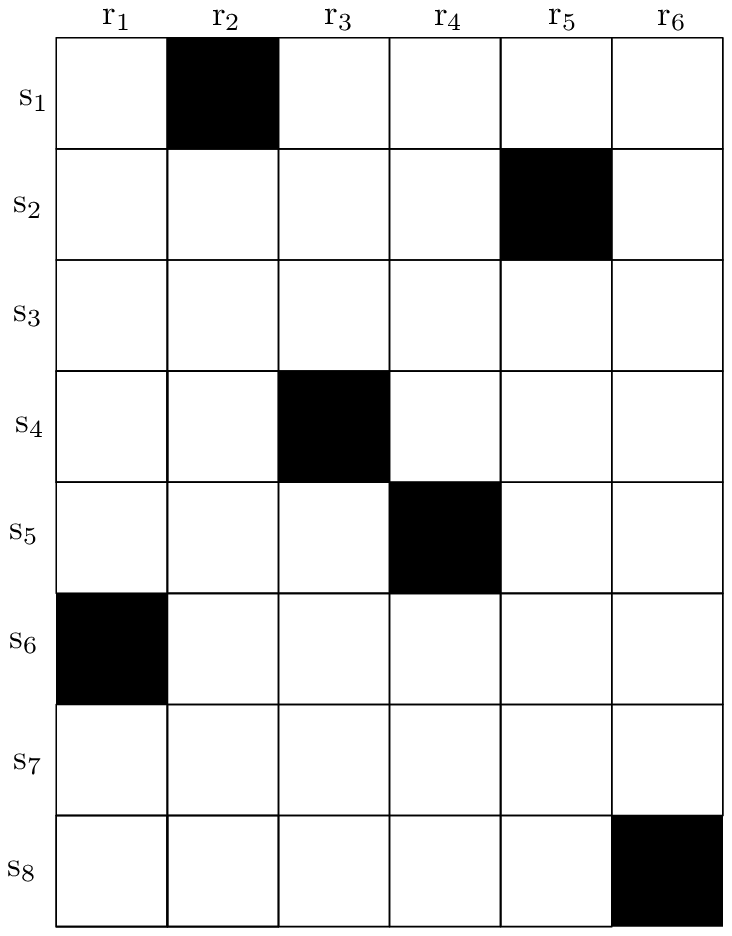}}
  \hspace{0.1in}
  \subfigure[]{
     \includegraphics[width=1.3in, height = 1.5in]{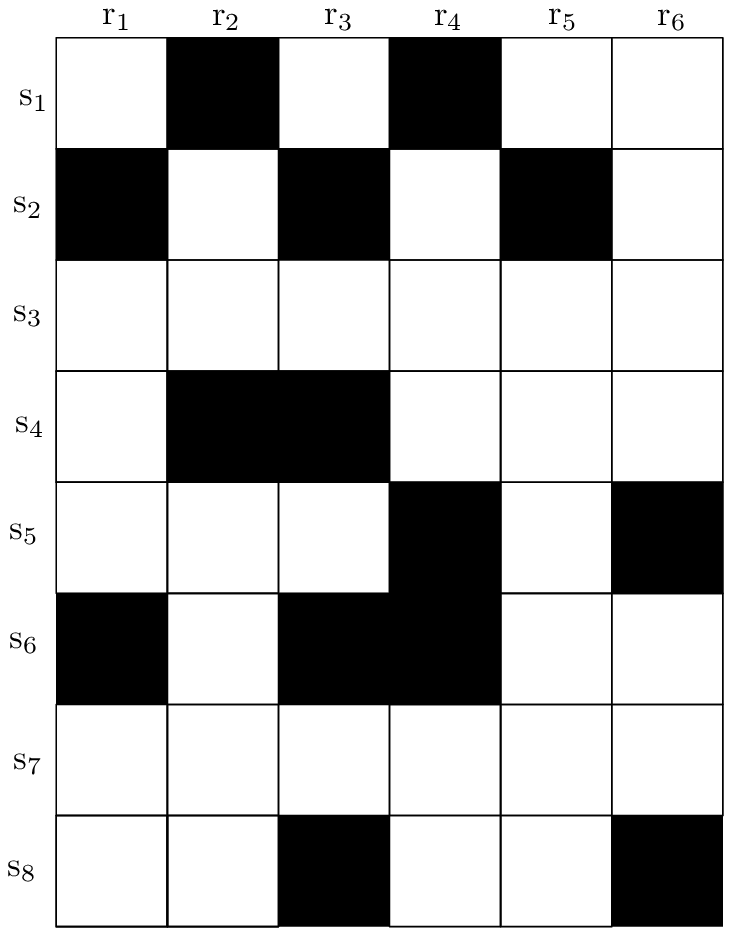}}
  \caption{Example patterns of selecting
           (a) one relay or (b) multiple relays assisting
           at each {relay-aided} subcarrier,
           where a black block at the $i$-th row and $j$-th column indicates
           the $j$-th relay ${\rm r}_j$ assists the transmission at the $i$-th subcarrier ${\rm s}_i$.
           }
  \label{fig:RelayPattern}
\end{figure}

So far, the majority of proposed RA algorithms, as the aforementioned ones,
restrict that at most one relay can assist the source at every {relay-aided} subcarrier.
In fact, when there are multiple relays available, allowing not just one,
but {\it each} of them to be eligible for assisting
can exploit all degrees of freedom in the system to improve performance.
For illustration purposes, example patterns of selecting single or multiple
assisting relays are shown in Figure \ref{fig:RelayPattern}.
For example, it has been shown in \cite{Vardhe10} that the sum power can be reduced
in a multi-user OFDM system if multiple DF relays assist the transmission
at every {relay-aided} subcarrier.

In this paper, we address the sum rate maximized RA problem in an OFDM system aided by
multiple DF relays subject to the individual (per device, i.e. a source or a relay)
sum (over all subcarriers) power constraints\footnote{
In this paper, the word ``individual" will mean per device and
``sum" will refer to a summation over all subcarriers, unless otherwise stated.}
of the source and the relays.
In particular, one or several relays may cooperate with the source to transmit
at every {relay-aided} subcarrier.
{When the sum power consumed by the source and that by every relay are fixed,
the optimum RA to this problem leads to the maximum energy efficiency for the system under consideration,
because the total energy consumed by the source and the relays
for transmitting per information bit is minimized.}
Specifically, our contributions lie in the following aspects:
\begin{itemize}
\item
{when a big number of subcarriers is used, it is shown that the duality gap of
the RA problem is equal to zero, based on the same idea first proposed in \cite{Yu06}.
Assuming the number of subcarriers is sufficiently large,
we develop a duality based algorithm which finds a globally optimum RA for the considered problem.
Most interestingly, the {\it sensitivity analysis} in convex optimization theory
is used to derive a {\it closed-form} optimum solution to a related convex optimization problem,
for which the method based on the Karush-Kuhn-Tucker (KKT) conditions is not applicable.}
\item
we develop a coordinate-ascent based iterative algorithm,
which finds a suboptimum RA but is always applicable
regardless of the duality gap of the RA problem.
Specifically, this algorithm produces a successive set of RAs with nondecreasing sum rate until convergence.
\end{itemize}

The remainder of this paper is organized as follows.
In the next section, the OFDM system under consideration is described and the RA problem is formulated.
Then, the duality based algorithm and the iterative algorithm are developed in Sections III and IV, respectively.
In Section V, the effectiveness of the proposed algorithms is illustrated by numerical experiments.
Finally, some conclusions wrap up this paper in Section VI.

\section{System description and RA problem formulation}\label{sec:system}

\subsection{System description}

We consider the OFDM transmission from a source to a destination
aided by $N$ DF relays collected in the set $\Rset=\{\ri | i=1,\cdots,N\}$.
All links are assumed to be frequency selective, and OFDM with properly
designed cyclic prefix is used to transform every link into $K$
parallel channels, each at a different subcarrier facing flat fading.
At every subcarrier, the transmission of a symbol is in either the direct mode,
or the {relay-aided} mode spanning across two equal-duration
time slots, namely the broadcasting slot and the relaying slot.
{We assume the destination decodes the signal samples received at each subcarrier
separately from those received at any other subcarrier.}

We make the following assumptions about the RA in the system.
First,
the RA is determined by an algorithm running at a central controller,
which knows precisely the noise power at each node,
as well as the channel coefficients at every subcarrier from the source to every $\ri$,
from the source to the destination, and from every $\ri$ to the destination, respectively.
Second,
all channels remain unchanged within
a sufficiently long duration, over which RA can be carried out accordingly.
Third,
the RA information can be reliably disseminated to
the source, every relay, and the destination.

\begin{table}[!t]
  \centering
  \caption{Channel coefficient at subcarrier $k$ between any two of the source, $\ri$, and the destination.}\label{tab:chcoeff}
  \begin{tabular}{|c|c|c|c|}
     \hline
       source to destination  &  source to $\ri$  &  $\ri$ to destination \\
     \hline
       $\Cstdk$ & $\Cstik$ & $\Citdk$ \\
     \hline
  \end{tabular}
\end{table}

\begin{figure}[!br]
  \centering
  \subfigure[In the broadcasting slot]{
     \includegraphics[width=1.5in]{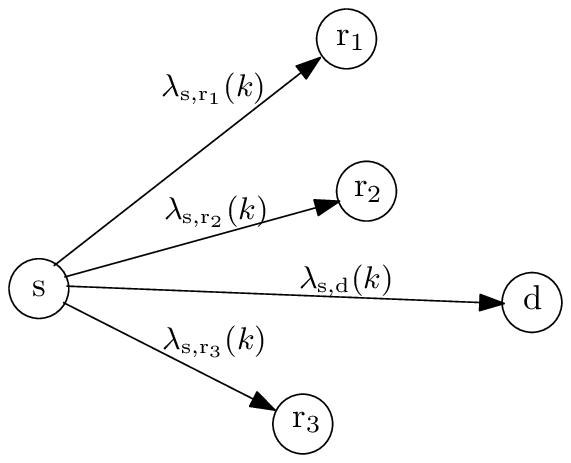}}
  \hspace{0.1in}
  \subfigure[In the relaying slot]{
     \includegraphics[width=1.5in]{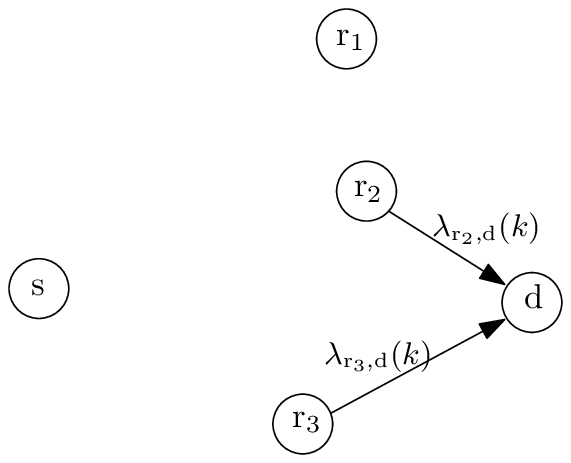}}
  \caption{Illustration of a {relay-aided} transmission where r$_2$ and r$_3$ assist relaying at subcarrier $k$.}
  \label{fig:RelayTx}
\end{figure}

Let's consider the transmission of a unit-variance symbol ${\theta}$ at subcarrier $k$.
The coefficient of the channel between any two of the source, $\ri$, and the destination,
are notated according to Table \ref{tab:chcoeff}.
We first describe the transmission in the {relay-aided} mode.
The source first emits in the broadcasting slot the symbol $\sqrt{\Ps\Psk}{{\theta}}$
as illustrated in Figure \ref{fig:RelayTx}.a,
where $\Ps$ and $\Psk$ represent the source sum power and
the fraction of that sum power allocated to the transmission at subcarrier $k$, respectively.
At the end of this slot,
both the destination and the relays receive the source signal.
The signal samples received at the destination and $\ri$ can be expressed respectively by
\begin{equation}
    \YdkB = \sqrt{\Ps\Psk}\Cstdk{{\theta}} + \NdkB    \label{eq:sample-broadcasting}
\end{equation}
and
\begin{equation}
    \YikB = \sqrt{\Ps\Psk}\Cstik{{\theta}} + \NikB,
\end{equation}
where $\NdkB$ and $\NikB$ represent the corruption of the additive white Gaussian noise (AWGN)
at the destination and $\ri$, respectively.
We assume $\forall\,i$, $\NikB$ is a zero-mean circularly Gaussian random variable
with the variance $\NVar$. The signal to noise ratio (SNR) at $\ri$ can be computed as $\Psk\Gstik$,
where $\Gstik=\frac{\Ps|\Cstik|^2}{\NVar}$ represents the normalized channel power gain
from the source to $\ri$.

For subcarrier $k$, we define $\SRsetk$ as the set containing all relays
sorted in the increasing order of $\Gstik$, $\srk{i}$ as the $i$-th relay in $\SRsetk$,
and $\SRsetkitoN{i}$ as the set containing relays in $\SRsetk$ with indices from $i$ to $N$.
We assume the $\ri$ that has the minimum $\Gstik$ among all assisting relays
for subcarrier $k$ is $\srk{\bk}$ {where $\bk$ represents the index of that $\ri$ in $\SRsetk$},
and all $\ri$'s in $\SRsetkitoN{\bk}$ decode the received samples to recover ${\theta}$.

{After the recovery of ${\theta}$,
all relays in $\SRsetkitoN{\bk}$ transmit simultaneously to the destination in the relaying slot,
which in effect establishes a distributed multiple input and single output (MISO) transmission
link as illustrated in Figure \ref{fig:RelayTx}.b.
Specifically, $\ri\in\SRsetkitoN{\bk}$ transmits $w_{\ri}{\theta}$,
where $w_{\ri}$ is the complex weight for transmit beamforming
and satisfies $|w_{\ri}|^2 = \Pri\Prik$.
$\Pri$ and $\Prik$ represent the sum power of $\ri$ and
the fraction of that sum power allocated to subcarrier $k$, respectively.
To have the relays' signals add coherently when received at the destination,
$w_{\ri} = \sqrt{\Pri\Prik}e^{-j\arg{(\Citdk)}}$ is used,
where $\arg{(\Citdk)}$ stands for the phase of $\Citdk$.
It should be noted that this transmission protocol enables a flexible use
of all relays opportunistically through a general form of {\it adaptive} transmit beamforming,
in that $\bk$ and $\{\Prik | \ri\in\SRsetkitoN{\bk}\}$ are determined by the RA algorithm
depending on channel state information, as will be developed later.}
At the end of the relaying slot, the signal sample received at the destination is denoted by
\begin{equation}
    \YdkR = \sum_{\ri\in\SRsetkitoN{\bk}}\sqrt{\Pri\Prik}|\Citdk|{\theta} + \NdkR,  \label{eq:sample-relaying}
\end{equation}
where $\NdkR$ represents the AWGN corruption at the destination.
We assume $\NdkB$ and $\NdkR$ are independent
zero-mean circularly Gaussian random variables with the same variance $\NVar$.

{Finally, $\YdkB$ and $\YdkR$ are processed at the destination.
From \eqref{eq:sample-broadcasting} and \eqref{eq:sample-relaying},
it can be seen that the relay-aided transmission at subcarrier $k$ is in effect over a channel
with a single input and two outputs.
To achieve the capacity of this channel, the maximum ratio combining (MRC) should be used,
i.e., the destination combines $\YdkB$ and $\YdkR$ to construct a decision variable
\begin{align}
    c_k = & \sqrt{\Ps\Psk}(\Cstdk)^*\YdkB +                                                          \nonumber \\
          &  \hspace{2cm}   \big(\sum_{\ri\in\SRsetkitoN{\bk}}\sqrt{\Pri\Prik}|\Citdk|\big)^*\YdkR,
\end{align}
which is then decoded (please refer to page 179 of \cite{Fund-WCOM} for more details).}
After mathematical arrangements, the SNR for decoding $c_k$ is derived as
\begin{align}
    \SNRmrck = \Psk\Gstdk + \bigg(\sum_{\ri\in\SRsetkitoN{\bk}}\sqrt{\Prik\Gitdk}\bigg)^2    \label{eq:SNRmrck} 
\end{align}
where $\Gstdk=\frac{\Ps|\Cstdk|^2}{\NVar}$ and $\Gitdk=\frac{\Pri|\Citdk|^2}{\NVar}$
represent the normalized channel power gains from the source to the destination
and from $\ri$ to the destination, respectively.

{To ensure the reliable recovery of {$\theta$}
at every relay in $\SRsetkitoN{\bk}$, the source transmission rate at subcarrier $k$
should not be higher than $\log_2(1 + \Psk\Gstxk{\srk{\bk}})$ bits/two-slots (bpts).
Moreover, the source transmission rate should not be higher than $\log_2(1 + \SNRmrck)$ bpts
to ensure reliable decoding of $c_k$ at the destination.
Therefore, the source transmission rate at subcarrier $k$
in the {relay-aided} mode should be \cite{Laneman04}
\begin{align}
   \RkR = & \min\big(\log_2(1 + \Psk\Gstxk{\srk{\bk}}), \log_2(1 + \SNRmrck)\big)   \nonumber                \\
        = &\log_2\big(1 + \min(\SNRmrck,\,\Psk\Gstxk{\srk{\bk}}) \big)  \quad {\rm bpts.}\label{eq:RkR}
\end{align}}

When $\bk$ is fixed, $\RkR$ is a concave function of $\{\Psk,\Prik | \forall\,\ri\in\Rset\}$
due to the following reasons.
First, $\SNRmrck$ and $\Psk\Gstxk{\srk{\bk}}$ are
both concave functions of $\{\Psk,\Prik | \forall\,\ri\in\Rset\}$,
and therefore $\min\{\SNRmrck,\,\Psk\Gstxk{\srk{\bk}}\}$
is a concave function of $\{\Psk,\Prik | \forall\,\ri\in\Rset\}$,
because the minimum of two concave functions is still a concave function.
Second, $\log_2(1+x)$ is an increasing and concave function of $x$.
Therefore, $\RkR$, as the composition of $\log_2(1+x)$ and $x=\min\{\SNRmrck,\,\Psk\Gstxk{\srk{\bk}}\}$,
is a concave function of $\{\Psk,\Prik | \forall\,\ri\in\Rset\}$ when $\bk$ is fixed,
because the composition of an increasing concave function and a concave function
is still a concave function \cite{Convex-opt}.

{As for the direct transmission mode at subcarrier $k$,
we consider a more efficient protocol compared to restricting the source to transmit only in the broadcasting slot
as in the related works \cite{Laneman04,Gui08,Li08,Vandendorpe08-1}.}
Specifically, the source emits two independent symbols in the two slots, respectively,
{and only the destination decodes the corresponding two received signal samples}.
We assume the AWGN corruptions for the two received samples
are independent zero-mean circularly Gaussian distributed with variance $\NVar$,
and the source uses the power $\Ps\Psk$ in total to transmit the two symbols.
The maximum source transmission rate at subcarrier $k$ in the direct mode can be derived as
\begin{align}
    \RkD = 2\log_2 \bigg(1 + \Psk\frac{\Gstdk}{2}\bigg)   \quad   {\rm bpts},   \label{eq:RkD}
\end{align}
and it is achieved when the source uses the power $\frac{\Ps\Psk}{2}$
to transmit each symbol.

{Note that we assume the same subcarrier is used by the source and the relays for transmitting a symbol in the relay-aided mode.
In fact, optimized subcarrier pairing could also be implemented,
which would further increase the degrees of freedom for optimization \cite{Li08}.
However, it would be more difficult to solve the RA problem,
and that is why in the current work, the RA algorithms are designed
under the assumption of nonoptimized subcarrier pairing.
We will nevertheless use the insights gained here to guide future work, which will also
consider subcarrier pairing.}

\subsection{Formulation of the RA problem}

We consider the RA problem to maximize the sum rate,
by optimizing the transmission mode and source power allocation for each subcarrier,
as well as assisting relays and power allocation to them for every {relay-aided} subcarrier.
To formulate this RA problem, we define a binary variable $\tk$
which indicates the transmission at subcarrier $k$
is in the relay-aided mode (resp. the direct mode) if $\tk=1$ (resp. $\tk=0$).
Mathematically, the RA problem is formulated as
\begin{align}
\max            &\hspace{0.25cm}   \sum_{k=1}^K \big(\tk\RkR + (1-\tk)\RkD\big)                                                          \nonumber  \\
\mathrm{s.t.}   &\hspace{0.25cm}   \sum_{k=1}^K \Psk \leq 1,   \hspace{0.6cm}   \sum_{k=1}^K \Prik \leq 1, \forall\,\ri\in\Rset,    \label{eq:prob} \\
                &\hspace{0.25cm}   \Psk\geq0, \forall\,k,      \hspace{0.6cm}   \Prik\geq0, \forall\,k, \forall\,\ri\in\Rset,            \nonumber  \\
                &\hspace{0.25cm}   \tk\in\{0,1\}, \forall\,k,  \hspace{0.3cm} \bk \in \{1,\cdots,N\},\forall k,                          \nonumber
\end{align}
where $\{\tk,\bk,\Psk,\Prik | \forall\,\ri\in\Rset,\forall\,k\}$ are to be optimized by the RA algorithm.

To facilitate analysis in the following sections,
\eqref{eq:prob} can be formulated into the following equivalent form
\begin{align}
\max_{\xvec}        &\hspace{0.25cm}   f(\xvec)                         \nonumber          \\
\mathrm{s.t.}       &\hspace{0.25cm}   \xvec\in\xDom,                   \label{eq:oriprob} \\
                    &\hspace{0.25cm}    g(\xvec) \preccurlyeq \onevec,  \nonumber  
\end{align}
where $\xvec$ represents the vector stacking all optimization variables,
$\xDom$ stands for the definition domain of $\xvec$,
$f(\xvec)$ denotes the sum rate,
$g(\xvec) = [g_1(\xvec),\cdots,g_{N+1}(\xvec)]^T$ where
$g_1(\xvec)=\sum_{k=1}^K\Psk$ and $g_{i+1}(\xvec)=\sum_{k=1}^K\Prik$ ($i=1,\cdots,N$)
stacks the sum power of the source and those of the relays,
$\onevec$ represents an $(N+1)\times1$ vector with every entry equal to $1$,
and $\preccurlyeq$ denotes the entrywise ``smaller than" inequality.

\section{The duality based RA algorithm}\label{sec:duality}

The Lagrangian for \eqref{eq:oriprob} is defined as
\begin{align}
L(\xvec,\muvec) = f(\xvec) + \muvec^T \big(\onevec - g(\xvec) \big),   \label{eq:defLag}
\end{align}
where $\muvec=[\muS,\mu_{{\rm r}_1}, \cdots, \mu_{{\rm r}_N}]^T$, with
$\muS$ and $\muri$ representing the dual variables related to the sum power constraints
of the source and $\ri$, respectively.
The dual function is defined as $d(\muvec) = \max_{\xvec\in\xDom}L(\xvec,\muvec)$,
and we denote a $\xvec$ that maximizes $L(\xvec,\muvec)$ as $\xvec_{\muvec}$.
If there exist multiple $\xvec$'s that maximize $L(\xvec,\muvec)$,
$d(\muvec)$ is not differentiable at $\muvec$,
and a subgradient of $d(\muvec)$ at $\muvec$ is equal to $\onevec - g(\xvec_{\muvec})$,
where any $\xvec$ that maximizes $L(\xvec,\muvec)$ can be chosen as $\xvec_{\muvec}$ \cite{Nonlinear-opt}.

It is important to note that the optimum objective value of \eqref{eq:oriprob},
denoted by $f^\star$, always satisfies $f^\star{\leq}d(\muvec)$ for any $\muvec\succcurlyeq\zerovec$,
where $\zerovec$ represents an $(N+1)\times1$ vector with every entry equal to $0$,
and $\succcurlyeq$ denotes the entrywise ``greater than" inequality \cite{Convex-opt}.
The duality gap is defined as $\min_{\muvec:\muvec\succcurlyeq\zerovec}{d(\muvec)}-f^\star$.
Note that $\xDom$ is not a convex set,
and therefore the duality gap of \eqref{eq:oriprob} might not be zero
since the Slater constraint qualification is not satisfied \cite{Nonlinear-opt}.
Interestingly, it is shown in Appendix \ref{sec:gap} based on the same idea proposed in \cite{Yu06} that,
when $K$ is sufficiently large \eqref{eq:oriprob} has zero duality gap.

Assuming sufficiently large $K$,
a duality based RA algorithm will be developed in this section
to find a globally optimum solution to \eqref{eq:oriprob}.
This algorithm relies on finding the optimum dual variable
$\optmuvec$ that minimizes $d(\muvec)$ based on the subgradient method,
i.e., updating $\muvec$ with $\muvec = [\muvec - \delta_q(\onevec - g(\xvec_{\muvec}))]^+$,
where $[\muvec]^+$ represents a vector built from $\muvec$ by only raising its negative entries to zero,
and $\delta_q$ represents the step size used in the $q$-th iteration,
until $g(\xvec_{\muvec})\preccurlyeq\onevec$ and $\muvec^T(\onevec-g(\xvec_{\muvec}))=0$ are fulfilled.
Then, $\xvec_{\optmuvec}$ is a globally optimum solution to \eqref{eq:oriprob}
as explained in Appendix A.
If $\delta_q$ satisfies the diminishing conditions
(i.e. $\lim_{q\rightarrow+\infty}\delta_q\rightarrow0$ and $\sum_{q=1}^{+\infty}\delta_q=\infty$),
$\muvec$ approaches $\optmuvec$ as $q$ increases \cite{Nonlinear-opt}.

In practice, it may take an unaffordably large number of iterations
before these optimality conditions are satisfied.
To address this issue, the iteration can be terminated when $g(\xvec_{\muvec})\preccurlyeq\onevec$
and $\muvec^T(\onevec-g(\xvec_{\muvec}))<\epsilon$,
where $\epsilon$ is a prescribed small positive value.
In this case,  $f(\xvec_{\muvec})\geq{f^\star-\epsilon}$ follows because of
\begin{align}
  f^\star - f(\xvec_{\muvec}) & \leq  d(\muvec) - f(\xvec_{\muvec})                    \nonumber  \\
                              & =   \muvec^T(\onevec-g(\xvec_{\muvec}))  < \epsilon    \nonumber,
\end{align}
which means that $\xvec_{\muvec}$ is a good approximation of a globally optimum solution
to \eqref{eq:oriprob}.

The overall duality based RA algorithm is summarized in {\bf Algorithm \ref{Alg:duality}}.
Specifically, $\muvec$ is initialized with $\onevec$ at the beginning,
and $\xvec_{\muvec}$ is found as the optimum solution to the Lagrangian maximization problem
\begin{align}
\max_{\xvec}     &\hspace{0.25cm}   L(\xvec,\muvec) =  \sum_{k=1}^K \big(\tk\RkR + (1-\tk)\RkD\big) +    \nonumber \\
                 &\hspace{2cm}         \muS\big(1-\sum_{k=1}^K \Psk\big) + \sum_{i=1}^N\muri\big(1-\sum_{k=1}^K\Prik\big),       \nonumber \\
\mathrm{s.t.}    &\hspace{0.25cm}   \Psk\geq0, \forall\,k,      \hspace{0.6cm}   \Prik\geq0, \forall\,k, \forall\,\ri\in\Rset,            \label{eq:MaxLag} \\
                 &\hspace{0.25cm}   \tk\in\{0,1\}, \forall\,k,  \hspace{0.3cm} \bk \in \{1,\cdots,N\},\forall k,                          \nonumber
\end{align}
with {\bf Algorithm \ref{Alg:MaxLag}} developed in Section \ref{sec:MaxLag}.
Furthermore, we choose $\delta_q=\frac{1 + Q_1}{q + Q_1}$ ($Q_1$ is a prescribed positive integer)
which satisfies the aforementioned diminishing conditions.

\begin{algorithm}
\caption{The duality based RA algorithm}\label{Alg:duality}
\begin{algorithmic}
{\small
\STATE  $q=1$, $\muvec = \onevec$;      \COMMENT{$\onevec$ is a vector with every entry equal to $1$}
\REPEAT
       \STATE $\muvec = [\muvec - \frac{1 + Q_1}{q + Q_1}(\onevec - g(\xvec_{\muvec}))]^+$
       \STATE $q = q+1$;
       \STATE Find $\xvec_{\muvec}$ with {\bf Algorithm \ref{Alg:MaxLag}};
\UNTIL{$g(\xvec_{\muvec})\preccurlyeq\onevec$ and $\muvec^T (\onevec-g(\xvec_{\muvec})) <\epsilon$}
\STATE $\xvec_{\muvec}$ produced at last is output as an optimum solution to \eqref{eq:oriprob}.
}
\end{algorithmic}
\end{algorithm}

\subsection{Algorithm to solve the Lagrangian maximization problem}\label{sec:MaxLag}

We can see that the optimum solution to \eqref{eq:MaxLag} can be found on a per subcarrier basis, i.e.,
$\{\tk(\muvec),\bk(\muvec),\Psk(\muvec),\Prik(\muvec) | \forall\,\ri\in\Rset\}$
representing the $\{\tk,\bk,\Psk,\Prik | \forall\,\ri\in\Rset\}$
contained in $\xvec_{\muvec}$ for subcarrier $k$, can be found as an optimum solution to
\begin{align}
\max            &\hspace{0.25cm}   L_k = \tk\RkR + (1-\tk)\RkD - \muS\Psk - \sum_{i=1}^N \muri\Prik        \nonumber \\
\mathrm{s.t.}   &\hspace{0.25cm}   \Psk\geq0,              \hspace{1.7cm} \Prik\geq0,\forall\,\ri\in\Rset,      \label{eq:MaxLag_k} \\
                &\hspace{0.25cm}   \bk \in \{1,\cdots,N\}, \hspace{0.5cm} \tk\in\{0,1\}.                    \nonumber
\end{align}

When $\tk$ and $\bk$ are fixed, \eqref{eq:MaxLag_k} is reduced to a convex optimization problem.
When $\tk=0$, the maximum $L_k$ is not influenced by $\bk$, while this not the case when $\tk=1$ and $\bk$ is fixed.
Based on the above analysis, an exhaustive-search based algorithm is designed to solve \eqref{eq:MaxLag_k}.
To facilitate algorithm design,
let's denote the maximum $L_k$ when $\tk=0$ by $\LkD$,
the maximum $L_k$ and the optimum $\{\Psk,\Prik | \forall\;\ri\in\Rset\}$
when $\tk=1$ and $\bk$ is fixed by $\LkR(\bk)$ and $\{\Psk(\bk),\Prik(\bk) | \forall\;\ri\in\Rset\}$, respectively.
Note that $\LkR(\bk)$ and $\{\Psk(\bk),\Prik(\bk) | \forall\;\ri\in\Rset\}$
can be evaluated with {\bf Algorithm \ref{Alg:MaxLag_k_tk_one}} developed in Section III.B.

It is important to note that $\LkR(\bk)\leq\LkD$ if $\Gsrbk\leq\Gstdk$,
because of
\begin{align}
\LkR(\bk) &= \log_2(1 + \min(\SNRmrck',\Psk(\bk)\Gsrbk)) + X_{\bk}    \nonumber \\
          &\leq \log_2(1 + \Psk(\bk)\Gstdk) + X_{\bk}                         \label{eq:inequal-bk} \\
          &\leq 2\log_2\bigg(1 + \Psk(\bk)\frac{\Gstdk}{2}\bigg) + X_{\bk}              \nonumber \\
          &\leq \LkD,                                                        \nonumber
\end{align}
where $X_{\bk} = -\muS\Psk(\bk) - \sum_{i=1}^N \muri\Prik(\bk)$,
and $\SNRmrck'$ represents the $\SNRmrck$
corresponding to $\SRsetkitoN{\bk}$ and $\{\Psk(\bk),\Prik(\bk) | \forall\;\ri\in\Rset\}$.

\begin{algorithm}
\caption{Algorithm to solve \eqref{eq:MaxLag} when $\muvec$ is fixed}\label{Alg:MaxLag}
\begin{algorithmic}
{\small
    \FOR{$k=1$ to $K$}
        \IF{$\Gstdk\geq\max_{\ri\in\Rset}\Gxtdk{\ri}$}
           \STATE $\tk(\muvec)=0$;
           \STATE $\Psk(\muvec)$ and $\Prik(\muvec),\forall\;\ri\in\Rset$ are evaluated by
                  \eqref{eq:optPs_tk_zero} and \eqref{eq:optPri_tk_zero}, respectively;
        \ELSE
           \STATE Compute $\LkD$ with \eqref{eq:LkD};
           \STATE Find $\ik$ as the minimum $i$ satisfying $\Gstxk{\srk{i}}>\Gstdk$;
           \FOR{$\bk=\ik$ to $N$}
                 \STATE Find $\{\Psk(\bk),\Prik(\bk) | \forall\;\ri\in\Rset\}$ with {\bf Algorithm \ref{Alg:MaxLag_k_tk_one}}
                        when $\tk=1$ and $\bk$ is fixed, then compute the $\LkR(\bk)$;
           \ENDFOR
           \IF{$\LkD\geq\max_{\bk\in\bkset}\LkR(\bk)$}
               \STATE $\tk(\muvec)=0$;
               \STATE $\Psk(\muvec)$ and $\Prik(\muvec),\forall\;\ri\in\Rset$ are evaluated by
                      \eqref{eq:optPs_tk_zero} and \eqref{eq:optPri_tk_zero}, respectively;
           \ELSE
               \STATE $\tk(\muvec)=1$;
               \STATE $\bk(\muvec)=\arg\max_{\bk\in\bkset}\LkR(\bk)$;
               \STATE $\{\Psk(\bk),\Prik(\bk) | \forall\;\ri\in\Rset\}$ when $\bk=\bk(\muvec)$
                      is assigned to $\{\Psk(\muvec),\Prik(\muvec) | \forall\;\ri\in\Rset\}$.
           \ENDIF
        \ENDIF
    \ENDFOR
}
\end{algorithmic}
\end{algorithm}

Based on the above analysis,
$\tk(\muvec)$, $\bk(\muvec)$, $\Psk(\muvec)$, and $\{\Prik(\muvec) | \forall\,\ri\in\Rset\}$
can be found with one of the following procedures:
\begin{enumerate}
\item
when $\max_{\ri}\Gstxk{\ri}\leq\Gstdk$,
$\tk(\muvec) = 0$, because $\LkD\geq\LkR(\bk)$ holds for every feasible value of $\bk$ according to \eqref{eq:inequal-bk}.
In this case, it can be derived according to the KKT conditions that
\begin{align}
&\Psk(\muvec)  = 2\bigg[\frac{\log_2(e)}{\muS} - \frac{1}{\Gstdk}\bigg]^+,   \label{eq:optPs_tk_zero}  \\
&\Prik(\muvec) = 0, \forall\;\ri\in\Rset,                            \label{eq:optPri_tk_zero} \\
&\LkD          = 2\log_2\left(1 + \Gstdk\bigg[\frac{\log_2(e)}{\muS} - \frac{1}{\Gstdk}\bigg]^+\right)    \label{eq:LkD}
\end{align}
\item
{when $\max_{\ri}\Gstxk{\ri}>\Gstdk$,
$\tk(\muvec)$ can be determined by an exhaustive-search based method,
i.e., if $\max_{\bk\in\{1,\cdots,N\}}\LkR(\bk)>\LkD$, $\tk(\muvec)=1$, otherwise $\tk(\muvec)=0$.
Let's denote $\ik$ as the minimum $i$ satisfying $\Gstxk{\srk{i}}>\Gstdk$.
Note that when $1 \leq \bk \leq \ik-1$, $\Gsrbk\leq\Gstdk$,
hence $\LkR(\bk)\leq\LkD$ holds for sure according to \eqref{eq:inequal-bk}.
This means that the comparison of $\LkD$ with $\max_{\bk\in\{1,\cdots,N\}}\LkR(\bk)$
is equivalent to comparing $\LkD$ with $\max_{\bk\in\bkset}\LkR(\bk)$ where $\bkset=\{\ik,\cdots,N\}$.
Based on this idea, $\LkD$ is evaluated with \eqref{eq:LkD},
and $\LkR(\bk)$ is computed for all values of $\bk\in\bkset$ with
{\bf Algorithm \ref{Alg:MaxLag_k_tk_one}}.
If $\LkD > \max_{\bk\in\bkset}\LkR(\bk)$, $\tk(\muvec) = 0$,
and $\Psk(\muvec)$ and $\{\Prik(\muvec) | \forall\;\ri\in\Rset\}$
are computed with \eqref{eq:optPs_tk_zero} and \eqref{eq:optPri_tk_zero}, respectively.
Otherwise,
$\tk(\muvec) = 1$, $\bk(\muvec)=\arg\max_{\bk\in\bkset}\LkR(\bk)$,
and the $\{\Psk(\bk),\Prik(\bk)|\forall\;\ri\in\Rset\}$ when $\bk=\bk(\muvec)$
is taken as $\{\Psk(\muvec),\Prik(\muvec) | \forall\;\ri\in\Rset\}$.}
\end{enumerate}

In summary, the overall algorithm of finding $\xvec_{\muvec}$ is summarized in {\bf Algorithm \ref{Alg:MaxLag}}.
We will proceed with developing {\bf Algorithm \ref{Alg:MaxLag_k_tk_one}} to
find $\LkR(\bk)$ and $\{\Psk(\bk),\Prik(\bk) | \forall\;\ri\in\Rset\}$ in the next section.

\subsection{Algorithm to solve \eqref{eq:MaxLag_k} when $\tk=1$ and $\bk$ is fixed} \label{sec:MaxLag_k_tk_one}

When $\tk=1$ and $\bk$ has a fixed value in $\bkset$, \eqref{eq:MaxLag_k} is equivalent to
\begin{align}
\max            &\;   L_k = \RkR - \muS\Psk - \sum_{i=1}^N \muri\Prik                \nonumber \\
                &\;    \quad =  \log_2(1 + \SNRk) - \muS\Psk - \sum_{i=1}^N \muri\Prik    \nonumber \\
\mathrm{s.t.}   &\;   \SNRk\leq\Psk\Gsrbk,                                                   \label{eq:MaxLag_k_tk_one} \\
                &\;   \SNRk\leq\Psk\Gstdk + \bigg(\sum_{\ri\in\SRsetkitoN{\bk}}\sqrt{\Prik\Gitdk}\bigg)^2,     \nonumber \\
                &\;   \SNRk \geq 0,                                                          \nonumber \\
                &\;   \Prik \geq 0,\forall\,\ri\in\Rset,         \nonumber
\end{align}
where $\SNRk$ is an intermediate optimization variable to guarantee the equivalence.

It can readily be shown that \eqref{eq:MaxLag_k_tk_one} is a convex optimization problem.
To solve it, one may formulate a set of equations based on the KKT conditions
and then solve them for $\{\Psk(\bk),\Prik(\bk) | \forall\;\ri\in\Rset\}$.
{\it
It is very important to note that this method is effective
only when the objective function and all the constraint functions are differentiable at the optimum solution. }
However, the second term in the right hand side of the second constraint in \eqref{eq:MaxLag_k_tk_one}
is not differentiable at $\Prik=0$, $\forall\,\ri\in\SRsetkitoN{\bk}$.
This means that if $\exists\;\ri\in\SRsetkitoN{\bk}$, $\Prik(\bk)=0$,
{\it which might happen as shown later},
the KKT conditions based method is not capable of finding that optimum solution.

To address this issue, we solve \eqref{eq:MaxLag_k_tk_one} based on the idea that
$\Psk(\bk)$ is the optimum solution to
\begin{align}
\max            &\hspace{0.25cm}   \log_2(1 + \SNRk) - \muS\Psk                 \nonumber \\
\mathrm{s.t.}   &\hspace{0.25cm}  \SNRk\leq\Psk\Gsrbk,                             \nonumber \\
                &\hspace{0.25cm}   \SNRk\leq\Psk\Gstdk + x,               \label{eq:prob1_xo} \\
                &\hspace{0.25cm}   \SNRk\geq0,                                      \nonumber
\end{align}
and $\{\Prik(\bk) | \forall\;\ri\in\Rset\}$ is the optimum solution to
\begin{align}
\max            &\hspace{0.25cm}   \sum_{i=1}^N(-\muri\Prik)              \nonumber \\
\mathrm{s.t.}   &\hspace{0.25cm}   \sum_{\ri\in\SRsetkitoN{\bk}}\sqrt{\Prik\Gitdk} = \sqrt{x},        \label{eq:prob2_xo} \\
                &\hspace{0.25cm}   \Prik\geq0,  \forall\,\ri\in\Rset,                   \nonumber
\end{align}
when $x=\xo$ with
\begin{align}
   \xo = \bigg(\sum_{\ri\in\SRsetkitoN{\bk}}\sqrt{\Prik(\bk)\Gitdk}\bigg)^2.   \nonumber
\end{align}

Specifically, $\xo$ is first determined, and then \eqref{eq:prob1_xo} and \eqref{eq:prob2_xo} with $x = \xo$
are solved to compute $\{\Psk(\bk),\Prik(\bk) | \forall\;\ri\in\Rset\}$. 
At first glance, this method seems confronted with a chicken-and-egg dilemma:
though $\{\Psk(\bk),\Prik(\bk) | \forall\;\ri\in\Rset\}$ can be computed
by solving \eqref{eq:prob1_xo} and \eqref{eq:prob2_xo} once $\xo$ is known,
it seems that $\{\Psk(\bk),\Prik(\bk) | \forall\;\ri\in\Rset\}$ needs to be known first
in order to compute $\xo$.
In fact, this dilemma can be elegantly circumvented by using 
the {\it sensitivity analysis} in convex optimization theory to first determine $\xo$ 
without knowing $\{\Psk(\bk),\Prik(\bk) | \forall\;\ri\in\Rset\}$, 
as elaborated in the following.

\subsubsection{Solutions to \eqref{eq:prob1_xo} and \eqref{eq:prob2_xo} given $x$}

Let's denote the optimum objective values of \eqref{eq:prob1_xo}
and \eqref{eq:prob2_xo} by $\optfone(x)$ and $\optftwo(x)$, respectively.
Obviously, \eqref{eq:prob1_xo} is a convex optimization problem.
Let's denote the optimum $\SNRk$ and the optimum dual variables associated with
the first and second constraints of \eqref{eq:prob1_xo} by $\SNRk(x)$, $\alphak(x)$, and $\betak(x)$, respectively.
According to the KKT conditions of \eqref{eq:prob1_xo},
\begin{align}
 \muS     = \Gsrbk\alphak(x) + \Gstdk\betak(x)        \label{eq:muScond}
\end{align}
and
\begin{align}
 \SNRk(x) =& \left[\frac{\log_2{e}}{\alphak(x)+\betak(x)} -1 \right]^+  \label{eq:SNRkcond}
\end{align}
should be satisfied.

As for \eqref{eq:prob2_xo}, it can readily be derived that
\begin{itemize}
\item
when $\forall\;\ri\in\SRsetkitoN{\bk}$, $\muri>0$, $\optftwo(x) = -\frac{x}{\Gbk}$,
and the optimum $\Prik$ to \eqref{eq:prob2_xo} is
\begin{align}
\Prik = \left\{\begin{array}{ll}
                  0             &    {\rm if} \; \ri\notin\SRsetkitoN{\bk},   \\
                  \frac{\Gxtdk{\ri}}{(\muri\Gbk)^2}x  &  {\rm if}\; \ri\in\SRsetkitoN{\bk}, \label{eq:optPrik_case2}
              \end{array}
       \right.
\end{align}
where $\Gbk = \sum_{\ri\in\SRsetkitoN{\bk}}\Gxtdk{\ri}/\muri$.
\item
when $\exists\;\ri\in\SRsetkitoN{\bk}$, $\muri=0$, $\optftwo(x) = 0$,
and the optimum $\Prik$ to \eqref{eq:prob2_xo} is equal to $0$
if $\ri\notin\SRsetkitoN{\bk}$ or if $\ri\in\SRsetkitoN{\bk}$ with $\muri>0$.
The optimum $\{\Prik | \forall\; \ri\in\SRsetkitoN{\bk},\muri=0\}$ is any set of
nonnegative values satisfying
\begin{align}
   \sum_{\ri:\ri\in\SRsetkitoN{\bk},\muri=0}\sqrt{\Prik\Gitdk} = \sqrt{x}.   \label{eq:optPrik_case3}
\end{align}

Moreover, it can readily be shown based on the Schwartz inequality that,
$\forall\; \ri:\ri\in\SRsetkitoN{\bk} {\;\rm and\;} \muri=0$,
the optimum $\Prik$ satisfying \eqref{eq:optPrik_case3} and minimizing the sum power of relays
is $\Prik=\frac{\Gitdk{x}^2}{(\sum_{\ri:\ri\in\SRsetkitoN{\bk},\muri=0}\Gitdk)^2}$.
\end{itemize}

\subsubsection{Finding $\xo$ based on the sensitivity analysis}

Let's denote the $L_k$ in \eqref{eq:MaxLag_k_tk_one}
computed with the optimum $\{\SNRk,\Psk\}$ to \eqref{eq:prob1_xo}
and the optimum $\{\Prik | \forall\;\ri\in\Rset\}$ to \eqref{eq:prob2_xo} by $\LkR(x,\bk)$ when $x\geq0$.
Obviously, $\LkR(x,\bk) = \optfone(x) + \optftwo(x)$ and $\LkR(x,\bk)\leq\LkR(\bk)$,
since $\LkR(\bk)$ is defined in Section III.A as the maximum $L_k$ for \eqref{eq:MaxLag_k_tk_one},
while $\LkR(x,\bk)$ is the $L_k$ computed with the above mentioned $\{\SNRk,\Psk\}$
and $\{\Prik | \forall\;\ri\in\Rset\}$ which are feasible for \eqref{eq:MaxLag_k_tk_one}.
When $x = \xo$, the $\Psk$ and $\{\Prik | \forall\;\ri\in\Rset\}$ used for computing $\LkR(x,\bk)$
are equal to $\Psk(\bk)$ and $\{\Prik(\bk) | \forall\;\ri\in\Rset\}$, respectively,
and therefore $\LkR(\xo,\bk) = \LkR(\bk)$.
This means that $\xo=\arg\max_{x:x\geq0}\LkR(x,\bk)$.

To determine $\xo$, let's consider $\LkR'(x,\bk)$ which represents
the first order derivative of $\LkR(x,\bk)$ with respect to $x$.
According to convex optimization theory, $\betak(x)$ represents the sensitivity of
$\optfone(x)$ with respect to $x$, i.e., $\optfone'(x)=\betak(x)$
(please refer to pages 249-253 in \cite{Convex-opt} for more details).
Therefore,
\begin{align}
\LkR'(x,\bk) =
\left\{\begin{array}{ll}
            \betak(x) - 1/\Gbk  &  {\rm if\;} \forall\,\ri\in\SRsetkitoN{\bk},\muri>0, \\
            \betak(x)           &  {\rm if\;} \exists\,\ri\in\SRsetkitoN{\bk},\muri=0.
       \end{array} \label{eq:deriveLk}
\right.
\end{align}

Based on the above analysis, the determination of $\xo$, $\betak(\xo)$, $\alphak(\xo)$
and $\Psk(\bk)$ falls into one of the following cases:

\begin{itemize}
\item
when $\forall\;\ri\in\SRsetkitoN{\bk}$, $\muri>0$ and $\frac{\muS}{\Gstdk}\leq\frac{1}{\Gbk}$, $\xo=0$.
This is because $\LkR(x,\bk)$ is a nonincreasing function of $x$
since $\betak(x)\in[0,\muS/\Gstdk]$ and $\LkR'(x,\bk)\leq0$ for any $x\geq0$.
In this case, the first constraint in \eqref{eq:prob1_xo} when $x=\xo$ is relaxed,
whereas the second one is saturated, since $\Gstdk<\Gsrbk$.
Therefore, $\alphak(\xo) = 0$, $\betak(\xo) = \frac{\muS}{\Gstdk}$,
$\SNRk(\xo)=\left[\frac{\Gstdk\log_2{e}}{\muS} - 1\right]^+$, and
\begin{align}
\Psk(\bk) = \frac{\SNRk(\xo)}{\Gstdk} = \left[\frac{\log_2{e}}{\muS} - \frac{1}{\Gstdk}\right]^+.   \label{eq:optPsk_case1}
\end{align}

\item
when $\forall\;\ri\in\SRsetkitoN{\bk}$, $\muri>0$ and $\frac{\muS}{\Gstdk}>\frac{1}{\Gbk}$,
$\betak(\xo)=\frac{1}{\Gbk}$ since $\LkR'(\xo,\bk)=0$ should be satisfied.
In this case, $\alphak(\xo) = \frac{\muS - \Gstdk/\Gbk}{\Gsrbk} > 0$ because \eqref{eq:muScond} is satisfied,
and $\SNRk(\xo)$ can be computed with \eqref{eq:SNRkcond}.
This means that both constraints in \eqref{eq:prob1_xo} when $x=\xo$ are saturated,
and therefore
\begin{align}
\Psk(\bk) =& \frac{\SNRk(\xo)}{\Gsrbk} \nonumber \\
          =& \left[\frac{\log_2{e}}{\muS + \DelG/\Gbk} - \frac{1}{\Gsrbk} \right]^+, \label{eq:optPsk_case2}
\end{align}
and
\begin{align}
\xo =& \Psk(\bk)\DelG     \nonumber \\
    =& \left[\frac{\log_2{e}}{\muS/\DelG + 1/\Gbk} - \frac{\DelG}{\Gsrbk} \right]^+,     \label{eq:optxo_case2}
\end{align}
where $\DelG = \Gsrbk - \Gstdk$.

\item
when $\exists\;\ri\in\SRsetkitoN{\bk}$, $\muri=0$,
$\betak(\xo)=0$ since $\LkR'(\xo,\bk)=0$ should be satisfied.
In this case, $\alphak(\xo) = \frac{\muS}{\Gsrbk}$,
and $\SNRk(\xo)=\left[\frac{\Gsrbk\log_2{e}}{\muS} - 1\right]^+$.
This means that the first constraint in \eqref{eq:prob1_xo} when $x=\xo$ is saturated,
whereas the second one is relaxed.
Therefore,
\begin{align}
\Psk(\bk) = \frac{\SNRk(\xo)}{\Gsrbk} = \left[\frac{\log_2{e}}{\muS} - \frac{1}{\Gsrbk}\right]^+, \label{eq:optPsk_case3}
\end{align}
and $\xo$ can be any value satisfying $\xo\geq\SNRk(\xo) - \Psk(\bk)\Gstdk = \xth$
where
\begin{align}
\xth = \DelG\left[\frac{\log_2{e}}{\muS} - \frac{1}{\Gsrbk}\right]^+.
\end{align}
\end{itemize}

\begin{algorithm}
\caption{Algorithm to solve \eqref{eq:MaxLag_k_tk_one} when $\muvec$ is fixed}\label{Alg:MaxLag_k_tk_one}
\begin{algorithmic}
{\small
        \IF{$\forall\;\ri\in\SRsetkitoN{\bk}$, $\muri>0$ and $\frac{\muS}{\Gstdk}\leq\frac{1}{\Gbk}$}
           \STATE $\Psk(\bk)$ is evaluated by \eqref{eq:optPsk_case1};
           \STATE $\Prik(\bk)=0$, $\forall\;\ri\in\Rset$;
        \ELSIF{$\forall\;\ri\in\SRsetkitoN{\bk}$, $\muri>0$ and $\frac{\muS}{\Gstdk}>\frac{1}{\Gbk}$}
           \STATE $\Psk(\bk)$ is evaluated by \eqref{eq:optPsk_case2};
           \STATE $\forall\;\ri\notin\SRsetkitoN{\bk}$, $\Prik(\bk)=0$;
           \STATE $\forall\;\ri\in\SRsetkitoN{\bk}$, $\Prik(\bk)$ is evaluated by \eqref{eq:optPrik_case2}
                  with $x$ equal to $\xo$ computed with \eqref{eq:optxo_case2};
        \ELSIF{$\exists\;\ri\in\SRsetkitoN{\bk}$, $\muri=0$}
           \STATE $\Psk(\bk)$ is evaluated by \eqref{eq:optPsk_case3},
           \STATE $\forall\;\ri\notin\SRsetkitoN{\bk}$, $\Prik(\bk)=0$;
           \STATE $\forall\;\ri\in\SRsetkitoN{\bk}\;{\rm with}\;\muri>0$, $\Prik(\bk)=0$;
           \STATE $\forall\;\ri\in\SRsetkitoN{\bk}\;{\rm with}\;\muri=0$, $\Prik(\bk)=\frac{\Gitdk{\xth}^2}{(\sum_{\ri:\ri\in\SRsetkitoN{\bk},\muri=0}\Gitdk)^2}$.
        \ENDIF
}
\end{algorithmic}
\end{algorithm}

After knowing $\xo$, we can find the optimum $\{\Prik | \forall\;\ri\in\Rset\}$ to \eqref{eq:prob2_xo} when $x=\xo$
as $\{\Prik(\bk) | \forall\;\ri\in\Rset\}$.
Note that in the third case $\xo$ can be any value no smaller than $\xth$,
and $\{\Prik(\bk) | \forall\; \ri\in\SRsetkitoN{\bk},\muri=0\}$ can be any set of
nonnegative values satisfying \eqref{eq:optPrik_case3} with $x=\xo$.
To improve the system energy efficiency,
we choose $\xo=\xth$, and $\forall\; \ri:\ri\in\SRsetkitoN{\bk} {\;\rm and\;} \muri=0$,
$\Prik(\bk)=\frac{\Gitdk{\xth}^2}{(\sum_{\ri:\ri\in\SRsetkitoN{\bk},\muri=0}\Gitdk)^2}$
to minimize the sum power of the relays.

The overall algorithm of finding $\{\Psk(\bk),\Prik(\bk) | \forall\;\ri\in\Rset\}$
is summarized in {\bf Algorithm \ref{Alg:MaxLag_k_tk_one}}.
{Based on the above analysis, it can be seen that
the $\SNRmrck$ corresponding to $\{\Psk(\bk),\Prik(\bk) | \forall\;\ri\in\Rset\}$ and $\bk$
must be equal to or smaller than $\Psk(\bk)\Gsrbk$.}

\section{The iterative RA algorithm}\label{sec:iterative}

In case \eqref{eq:prob} has a nonzero duality gap,
{\bf Algorithm \ref{Alg:duality}} fails to find a globally optimum solution.
To address this issue,
we will develop in this section a coordinate-ascent based iterative algorithm
which is suboptimum but always applicable regardless of the duality gap of \eqref{eq:prob}.

\subsection{The iterative RA algorithm}

First of all, it should be noted that $\RkR\leq\RkD$ always holds independently of
$\{\Psk,\Prik | \forall\,\ri\in\Rset\}$ when $\bk$ satisfies $\Gsrbk\leq\Gstdk$.
This means that when $\max_{\ri}\Gstik\leq\Gstdk$,
$\tk=0$ is the optimum independently of $\bk$ and $\{\Psk,\Prik | \forall\,\ri\in\Rset\}$,
i.e., the direct transmission mode should always be used for subcarrier $k$.

To simplify algorithm design,
we assume $\tk$ is fixed as $0$ for every $k\in\DCarset=\{k|\max_{\ri}\Gstik\leq\Gstdk\}$.
If $k\notin\DCarset$,
the optimum $\tk$ and $\bk$ must lie in the set $\{0,1\}\times\bkset$,
where $\times$ is the Cartesian product operator.
Furthermore,
when $\{\tk,\bk | \forall\,k\notin\DCarset\}$ is fixed,
\eqref{eq:prob} is reduced to a concave maximization problem,
which has zero duality gap since the Slater constraint qualification is satisfied.
Thus a globally optimum $\{\Psk,\Prik | \forall\,\ri\in\Rset,\forall\,k\}$ and the maximum sum rate
when $\{\tk,\bk | \forall\,k\notin\DCarset\}$ is fixed
can be found with a duality based algorithm as shown later.

Now the difficulty of solving \eqref{eq:prob} lies in finding
the optimum $\{\tk,\bk | \forall\,k\notin\DCarset\}$.
To this end, one may use an exhaustive-search based algorithm.
Specifically, this algorithm finds the maximum sum rate
for each possible $\{\tk,\bk | \forall\,k\notin\DCarset\}$ in $\prod_{k\notin\DCarset}(\{0,1\}\times\bkset$).
Then it chooses the best one as the optimum $\{\tk,\bk | \forall\,k\notin\DCarset\}$.
However, the complexity of exhaustive search might be unaffordable
for practical systems using a big number of subcarriers.

To address this issue, we develop a coordinate-ascent based iterative algorithm
which produces a successive set of RAs with nondecreasing sum rate until convergence.
In the following, a superscript $m$ added to a variable indicates that variable
is produced at the $m$-th iteration to facilitate description.
At the beginning,
$\forall\;k\notin\DCarset$, $\tk^1$ and $\bk^1$ are initialized as $1$ and $N$, respectively,
i.e. every subcarrier not in $\DCarset$ is set in the relay-aided mode
with only the relay having the highest source-relay channel gain enabled for assisting.
In the $m$-th iteration, $\{\Psk^m,\Prik^m | \forall\,\ri\in\Rset,\forall\,k\}$,
which is the optimum solution to \eqref{eq:prob} when $\tk=\tk^m$ and $\bk=\bk^m$, $\forall\;k\notin\DCarset$,
is first found with a duality based algorithm, namely {\bf Algorithm \ref{Alg:duality-iterative}}
developed in Section \ref{sec:iterative-optpow}.
This step can be interpreted as finding the optimum source/relay power allocation
when $\{\tk,\bk | \forall\,k\notin\DCarset\}$ is fixed as $\{\tk^{m},\bk^{m} | \forall\,k\notin\DCarset\}$.

\begin{algorithm}
{\small
\caption{The iterative RA algorithm}\label{Alg:iterative}
\begin{algorithmic}
\STATE  $\forall\;k\in\DCarset$, $\tk=0$; $\forall\;k\notin\DCarset$, $\tk^1=1$ and $\bk^1=N$;
\STATE  $m = m + 1$;
\REPEAT
       \STATE Find $\{\Psk^m,\Prik^m | \forall\,\ri\in\Rset,\forall\,k\}$ as the optimum solution to \eqref{eq:prob}
                    when $\tk=\tk^m$ and $\bk=\bk^m$, $\forall\;k\notin\DCarset$ with {\bf Algorithm \ref{Alg:duality-iterative}};
       \FOR{every $k\notin\DCarset$}
                 \IF{$\tk^m=0$}
                     \STATE $\tk^{m+1}=0$;
                 \ELSE
                     \STATE Compute $\SNRmrck^m(\bk^m)$, $\RkR^m(\bk^m)$, and $\RkD^m$;
                     \IF{$\RkR^m(\bk^m) < \RkD^m$}
                           \STATE  $\tk^{m+1}=0$;
                     \ELSE
                           \STATE  $\tk^{m+1}=1$;
                           \STATE  $\bk^{m+1}$ is set as the minimum $i$ satisfying $\SNRmrck^m(\bk^m)\leq\Psk^m\Gstxk{\srk{i}}$;
                     \ENDIF
                 \ENDIF
       \ENDFOR
       \STATE $m = m + 1;$
\UNTIL{$\forall\;k\notin\DCarset$, $\{\tk^{m+1}, \bk^{m+1}\}$ is equal to $\{\tk^{m}, \bk^{m}\}$}
\STATE  $\{\Psk^m,\Prik^m,\forall\,\ri\in\Rset,\forall\,k\}$ and $\{\tk^m,\bk^m,\forall\;k\notin\DCarset,\tk,\bk,\forall\;k\in\DCarset\}$
       produced in the last iteration are the suboptimum solution.
\end{algorithmic}}
\end{algorithm}

{Then, $\{\tk^{m+1},\bk^{m+1} | \forall\,k\notin\DCarset\}$ which maximizes the sum rate
is found when the power allocation is fixed as $\{\Psk^m,\Prik^m | \forall\,\ri\in\Rset,\forall\,k\}$.
This step can be interpreted as finding the optimum mode and assisting relays
when the source/relay power allocation is prescribed by $\{\Psk^m,\Prik^m | \forall\,\ri\in\Rset,\forall\,k\}$.
Note that this can be accomplished on a per subcarrier basis,
i.e., for every subcarrier $k\notin\DCarset$, $\tk^{m+1}$ and $\bk^{m+1}$
are found to maximize the rate
when the power allocation is fixed as $\{\Psk^m,\Prik^m | \forall\,\ri\in\Rset\}$.
In this case, the rate is denoted by $\RkD^m = 2\log_2(1 + \frac{\Psk^m}{2}\Gstdk)$ if the direct mode is used.
If the relay-aided mode with a given $\bk$ is used, the rate is denoted by
$\RkR^m(\bk) = \log_2(1 + \min(\SNRmrck^m(\bk),\Psk^m\Gstxk{\srk{\bk}}))$
where $\SNRmrck^m(\bk)$ represents the SNR of MRC at the destination,

Let's first consider the evaluation of $\tk^{m+1}$ and $\bk^{m+1}$ when $\tk^m=0$,
meaning that subcarrier $k$ was set in the direct mode
when $\{\Psk^m,\Prik^m | \forall\,\ri\in\Rset,\forall\,k\}$ was evaluated.
Suppose the relay-aided mode with any $\bk^{m+1}$ is now used instead,
the rate is reduced independently of $\bk^{m+1}$ because of
\begin{align}
\RkR^m(\bk^{m+1}) &= \log_2(1 + \min(\SNRmrck^m(\bk^{m+1}),\Psk^m\Gstxk{\srk{\bk^{m+1}}}))            \nonumber \\
                  &\leq \log_2(1 + \SNRmrck^m(\bk^{m+1}))                                            \nonumber \\
                  & =   \log_2(1 + \Psk^m\Gstdk) \leq \RkD^m,          \nonumber
\end{align}
where the equality in the third line is because of
$\forall\;\ri\in\Rset$, $\Prik^m=0$ as will be shown in Section \ref{sec:iterative-optpow}.
In order to maximize the rate, $\tk^{m+1}=0$, i.e. subcarrier $k$ remains in the direct mode.

Next, we consider the evaluation of $\tk^{m+1}$ and $\bk^{m+1}$ when $\tk^m=1$,
meaning that subcarrier $k$ was set in the relay-aided mode with $\bk=\bk^m$
when $\{\Psk^m,\Prik^m | \forall\,\ri\in\Rset,\forall\,k\}$ was evaluated.
As will be shown in Section \ref{sec:iterative-optpow},
$\SNRmrck^m(\bk^m)\leq\Psk^m\Gstxk{\srk{\bk^m}}$ always holds and thus $\RkR^m(\bk^m) = \log_2(1 + \SNRmrck^m(\bk^m))$.
It can readily be seen from \eqref{eq:SNRmrck} that
if $\bk>\bk^m$, $\SNRmrck^m(\bk) \leq \SNRmrck^m(\bk^m)$,
while $\SNRmrck^m(\bk) = \SNRmrck^m(\bk^m)$ if $\bk \leq \bk^m$,
since $\forall\;\ri\notin\SRsetkitoN{\bk^m}$, $\Prik^m=0$ as will be shown in Section \ref{sec:iterative-optpow}.
This means that $\forall\;\bk$, $\RkR^m(\bk)\leq\log_2(1 + \SNRmrck^m(\bk))\leq\RkR^m(\bk^m)$,
i.e., if subcarrier $k$ remains in the relay-aided mode,
the rate can not be increased no matter which value is assigned to $\bk$. 
In this case, $\tk^{m+1}$ and $\bk^{m+1}$ are determined
in one of the following cases:
\begin{itemize}
\item
if $\RkD^m > \RkR^m(\bk^m)$,
$\tk^{m+1}$ is set as 0 since the rate is increased if the direct mode is used;
\item
if $\RkD^m \leq \RkR^m(\bk^m)$, $\tk^{m+1}$ is still set as $1$.
$\bk^{m+1}$ is assigned as the smallest $\bk$ satisfying $\RkR^m(\bk) = \RkR^m(\bk^m)$.
Obviously, $\bk^{m+1}\leq\bk^m$,
and $\SRsetkitoN{\bk^{m+1}}$ is the biggest set of assisting relays that leads to the same rate as $\SRsetkitoN{\bk^{m}}$.
The motivation behind this assignment is twofold.
One is to guarantee the rate when $\bk=\bk^{m+1}$ is no smaller than when $\bk=\bk^m$.
The other is to increase the degrees of freedom
for optimizing the power allocation in the next iteration,     
since $\forall\;\ri: \ri\in\SRsetkitoN{\bk^{m+1}}\; {\rm and}\; \ri\notin\SRsetkitoN{\bk^{m}}$,
$\Prik^m=0$ always holds but $\Prik$ is free to be optimized in the $(m+1)$-th iteration.
It can easily be shown that $\bk^{m+1}$ is actually equal to
the minimum $i$ satisfying $\SNRmrck^m(\bk^m)\leq\Psk^m\Gstxk{\srk{i}}$.
Note that $\Gstxk{\srk{\bk^{m+1}}}\geq\frac{\SNRmrck^m(\bk^m)}{\Psk^m}\geq\Gstdk$,
which means that $\bk^{m+1}\in\bkset$.
\end{itemize}}

It can easily be seen that a successive set of RAs with nondecreasing sum rate
are produced as the iteration proceeds.
After the algorithm converges, the RA produced by the last iteration
is output as a suboptimum solution.
This RA is at least a locally optimum solution to \eqref{eq:prob}
(please refer to pages 160-162 of \cite{Nonlinear-opt} for more details).
In summary, the iterative algorithm is described in {\bf Algorithm \ref{Alg:iterative}}.

\subsection{Algorithm to solve \eqref{eq:prob} when $\tk$ and $\bk$ are fixed}\label{sec:iterative-optpow}

When $\tk=\tk^m$ and $\bk=\bk^m$, $\forall\;k\notin\DCarset$, and $\tk=0$, $\forall\;k\in\DCarset$,
\eqref{eq:prob} is equivalent to
\begin{align}
\max            &\;   \sum_{k=1,k\notin\DCarset}^K (\tk^m\RkR + (1-\tk^m)\RkD) + \sum_{k\in\DCarset}\RkD      \nonumber  \\
\mathrm{s.t.}   &\;   \sum_{k=1}^K \Psk \leq 1,  \quad \sum_{k=1}^K \Prik \leq 1,   \quad \forall\,\ri,            \label{eq:prob-iterative} \\
                &\;   \Psk\geq0, \forall\,k,    \quad \Prik\geq0, \forall\,k, \forall\,\ri\in\Rset,        \nonumber
\end{align}
which has zero duality gap as shown earlier,
and therefore can be solved with a duality based algorithm.
To this end, we define the dual variables $\nuS$ and $\nuri$ related to
the sum power constraints of the source and $\ri$, respectively.
Let's stack $\nuS$ and $\{\nuri | \forall\,\ri\in\Rset\}$ into a vector $\nuvec$,
and $\{\Psk,\Prik | \forall\,\ri\in\Rset,\forall\,k\}$ into a vector $\Pvec$.
Then, we define the Lagrangian as
\begin{align}
W(\Pvec,\nuvec) =& \sum_{k=1,k\notin\DCarset}^K (\tk^m\RkR + (1-\tk^m)\RkD) + \sum_{k\in\DCarset}\RkD     \nonumber \\
                 & + \sum_{i=1}^N\nuS\big(1-\sum_{k=1}^K\Psk\big) + \sum_{i=1}^N\nuri\big(1-\sum_{k=1}^K\Prik\big).  \nonumber
\end{align}

The duality based algorithm looks for the optimum $\nuvec$ with the subgradient based method.
In each iteration, the optimum $\Pvec$ that maximizes $W(\Pvec,\nuvec)$ subject to
the constraints $\Psk\geq0$, $\Prik\geq0$, $\forall\;k$, $\forall\;\ri\in\Rset$,
denoted by $\Pvec_{\nuvec}$, is found when $\nuvec$ is fixed.
We denote $\Psk$ and $\Prik$ contained in $\Pvec_{\nuvec}$ as $\Psk(\nuvec)$ and $\Prik(\nuvec)$, respectively.
Specifically, for subcarrier $k$ with $\tk^m=1$ and $\bk=\bk^m$,
the problem of finding $\{\Psk(\nuvec),\Prik(\nuvec) | \forall\;\ri\in\Rset\}$
is equivalent to \eqref{eq:MaxLag_k_tk_one} with $\bk=\bk^m$, $\muS=\nuS$, and $\muri=\nuri$, $\forall\;\ri\in\Rset$,
and therefore can be solved with {\bf Algorithm \ref{Alg:MaxLag_k_tk_one}}
after replacing $\{\bk,\muS,\muri | \forall\;\ri\in\Rset\}$ with $\{\bk^m,\nuS,\nuri | \forall\;\ri\in\Rset\}$.
{It is important to note that the $\SNRmrck$ corresponding to
$\{\Psk(\nuvec),\Prik(\nuvec) | \forall\;\ri\in\Rset\}$ and $\bk = \bk^m$
must be equal to or smaller than $\Psk(\nuvec)\Gstxk{\srk{\bk^m}}$,
according to the analysis at the end of Section III.B. }
For subcarrier $k$ with $\tk^m=0$ or in $\DCarset$, the problem of finding $\{\Psk(\nuvec),\Prik(\nuvec) | \forall\;\ri\in\Rset\}$
is equivalent to \eqref{eq:MaxLag_k} with $\tk=0$, $\muS=\nuS$, and $\muri=\nuri$, $\forall\;\ri\in\Rset$,
and therefore can be solved with \eqref{eq:optPs_tk_zero} and \eqref{eq:optPri_tk_zero}
after replacing $\muS$ with $\nuS$.
Note that $\forall\,\ri\in\Rset$, $\Prik(\nuvec)=0$ if $k\in\DCarset$ or $\tk^m=0$,
and $\forall\,\ri\notin\SRsetkitoN{\bk^m}$, $\Prik(\nuvec)=0$ if $\tk^m=1$.

\begin{algorithm}[!t]
{\small
\caption{The duality based RA algorithm to solve \eqref{eq:prob}
when $\tk=\tk^m$ and $\bk=\bk^m$, $\forall\;k\notin\DCarset$, and $\tk=0$, $\forall\;k\in\DCarset$}\label{Alg:duality-iterative}
\begin{algorithmic}
\STATE  $j=1$, $\nuvec = \onevec$;
\REPEAT
       \STATE $\nuS =  [\nuS  - \frac{1 + Q_2}{j+Q_2}(1 - \sum_{k=1}^K\Psk(\nuvec))]^+$;
       \STATE $\nuri = [\nuri - \frac{1 + Q_2}{j+Q_2}(1 - \sum_{k=1}^K\Prik(\nuvec))]^+$, $\forall\;\ri\in\Rset$;
       \STATE $j = j+1$;

       \FOR{$k=1$ to $N$}
           \IF{$k\in\DCarset$ or $k\notin\DCarset$ with $\tk^m=0$}
                \STATE  $\Psk(\nuvec)$ and $\Prik(\nuvec)$, $\forall\;\ri$ are found
                        with \eqref{eq:optPs_tk_zero} and \eqref{eq:optPri_tk_zero}
                        after replacing $\muS$ with $\nuS$;
           \ELSIF{$k\notin\DCarset$ with $\tk^m=1$}
                \STATE  $\Psk(\nuvec)$ and $\Prik(\nuvec)$, $\forall\;\ri$ are found with
                        {\bf Algorithm \ref{Alg:MaxLag_k_tk_one}} after replacing
                        $\{\bk,\muS,\muri | \forall\;\ri\in\Rset\}$ with $\{\bk^m,\nuS,\nuri | \forall\;\ri\in\Rset\}$;

           \ENDIF
       \ENDFOR
\UNTIL{$\Pvec_{\nuvec}$ is feasible for \eqref{eq:prob-iterative} and \\ $\nuS(1-\displaystyle{\sum_{k=1}^K}\Psk(\nuvec))+\sum_{\ri\in\Rset}\nuri(1-\displaystyle{\sum_{k=1}^K}\Prik(\nuvec))<\epsilon$ }
\STATE $\Psk(\nuvec)$, and $\Prik(\nuvec)$, $\forall\,\ri\in\Rset$
       produced in the last iteration are the optimum solution to \eqref{eq:prob}
       when $\tk=\tk^m$ and $\bk=\bk^m$, $\forall\;k\notin\DCarset$, and $\tk=0$, $\forall\;\ri\in\DCarset$.
\end{algorithmic}}
\end{algorithm}

The overall algorithm is summarized as {\bf Algorithm \ref{Alg:duality-iterative}}.
The step size and the termination condition are chosen in the same way as in {\bf Algorithm \ref{Alg:duality}}.
{Specifically, we choose $\delta_j=\frac{1 + Q_2}{j + Q_2}$ ($Q_2$ is a prescribed positive integer)
which satisfies the diminishing conditions as proposed in \cite{Nonlinear-opt}.}
Based on the analysis in Section \ref{sec:duality},
the finally produced $\Pvec_{\nuvec}$ corresponds to a sum rate no smaller than $z^\star-\epsilon$
($z^\star$ represents the maximum objective of \eqref{eq:prob-iterative}).

\section{Numerical experiments}

\begin{figure}
  \centering
     \includegraphics[width=3.5in]{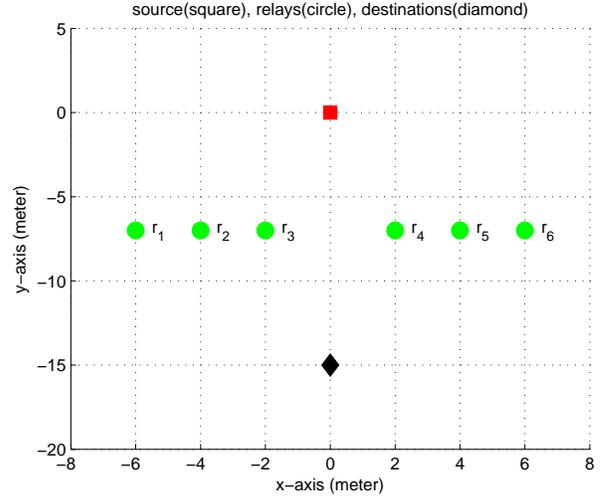}
  \caption{A DF relays aided OFDM transmission system considered in numerical experiments.}
  \label{fig:Systeup}
\end{figure}

For illustration purposes, we consider an OFDM transmission system
aided by $6$ DF relays shown in Figure \ref{fig:Systeup}.
The source is located at the origin, the destination is located at the coordinate $(0,-15)$,
and $\ri$, $i=1,\cdots,6$, is located at the coordinates $(-6, -7)$, $(-4, -7)$, $(-2, -7)$,
$(2, -7)$, $(4, -7)$, and $(6, -7)$, respectively.
Note that all the aforementioned coordinate-related values have the unit of meter.
The parameters are set as $\NVar=-50$ dBW, $\epsilon=0.1$, $Q_1=Q_2=50$,
{and the duration of one time slot is 1 millisecond}.
In numerical experiments, we found that the convergence of {\bf Algorithm \ref{Alg:duality}}
was always observed when $K\geq256$. Thus we choose $K=256$.

\begin{figure}
  \centering
     \includegraphics[width=3.5in]{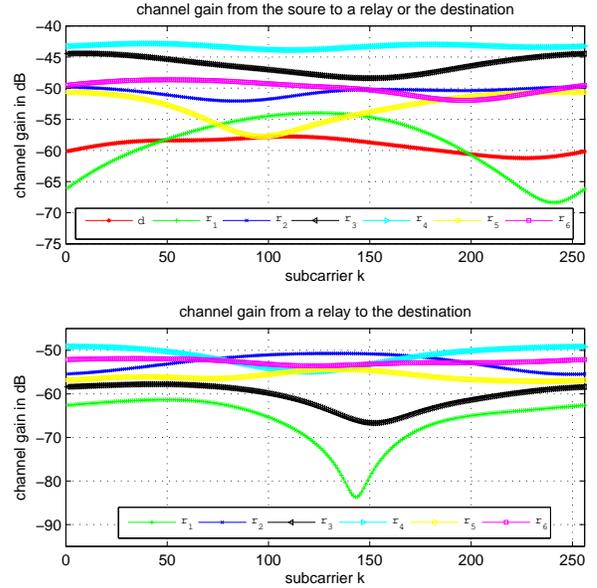}
  \caption{A random channel gain realization.}
  \label{fig:oneGain}
\end{figure}

We generate every channel according to the following assumptions.
First, it is modeled as a $6$-tap delay line,
and the average received power at a distance of $d$ is equal to $\xi{d}^{-3}$,
where $\xi$ represents a log-normal shadowing effect of $1$ dB
(i.e., $10\log_{10}(\xi)$ is Gaussian distributed with zero mean and variance equal to $1$ dB).
Second, we assume the amplitude of the $i$-th tap is a circularly symmetric
complex Gaussian random variable with zero mean and variance as
$\sigma_i^2 = \sigma_0^2{e}^{-3i}$, $i=0,1,\cdots,5$.

In order to illustrate the effectiveness of the proposed algorithms,
we consider a heuristic RA algorithm against which the proposed RA algorithms are compared.
{This heuristic algorithm} allocates the source sum power uniformly to all subcarriers, i.e., $\Psk=\frac{1}{K}$.
Then, for every $\ri$,
$\Omega_{\ri} = \{k|k\notin\DCarset,\Gstik=\max_{{\rm r}_j\in\Rset}\Gstxk{{\rm r}_j}\}$
which contains every subcarrier not belonging to $\DCarset$ and at which $\ri$
has the maximum source to relay channel gain, is first found.
The sum power of $\ri$ is uniformly allocated to all subcarriers in $\Omega_{\ri}$
(i.e., if $k\notin\Omega_{\ri}$, $\Prik=0$, otherwise
$\Prik=\frac{1}{|\Omega_{\ri}|}$ where $|\Omega_{\ri}|$ denotes the number of subcarriers in $\Omega_{\ri}$).
For every subcarrier $k\in\DCarset$, the direct transmission mode is used.
For every subcarrier $k\notin\DCarset$, $\RkD$ is first computed,
and then the $\RkR$ when a single $\ri$ with the maximum $\Gstik$ assists relaying is computed.
If $\RkD$ is equal to or greater than the computed $\RkR$, the direct mode is used.
Otherwise, the relay mode is used, and only the $\ri$ used for computing $\RkR$ assists relaying.

\begin{figure}
  \centering
     \includegraphics[width=3.5in]{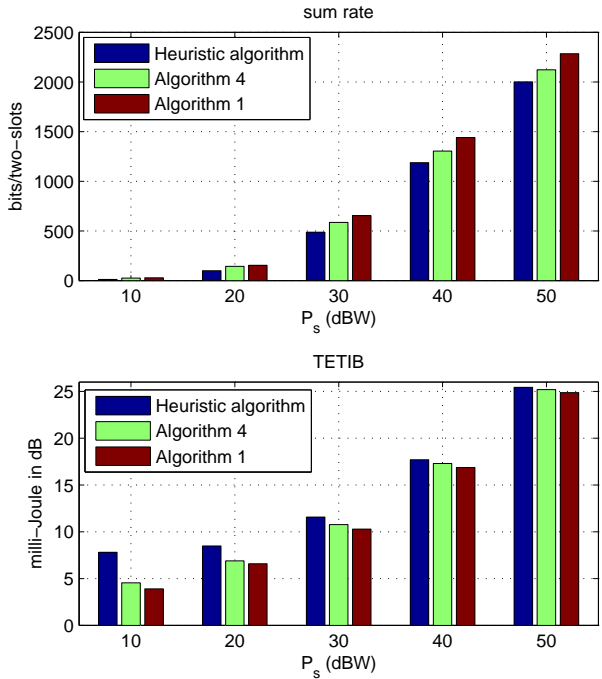}
  \caption{For the RA evaluated by each algorithm, the sum rate and the TETIB
           when $\Ps=\Pri$, $\forall\;\ri\in\Rset$, and $\Ps$ varies from $10$ to $50$ dBW.}  \label{fig:rate}
\end{figure}

\begin{figure}
  \centering
     \includegraphics[width=3.5in]{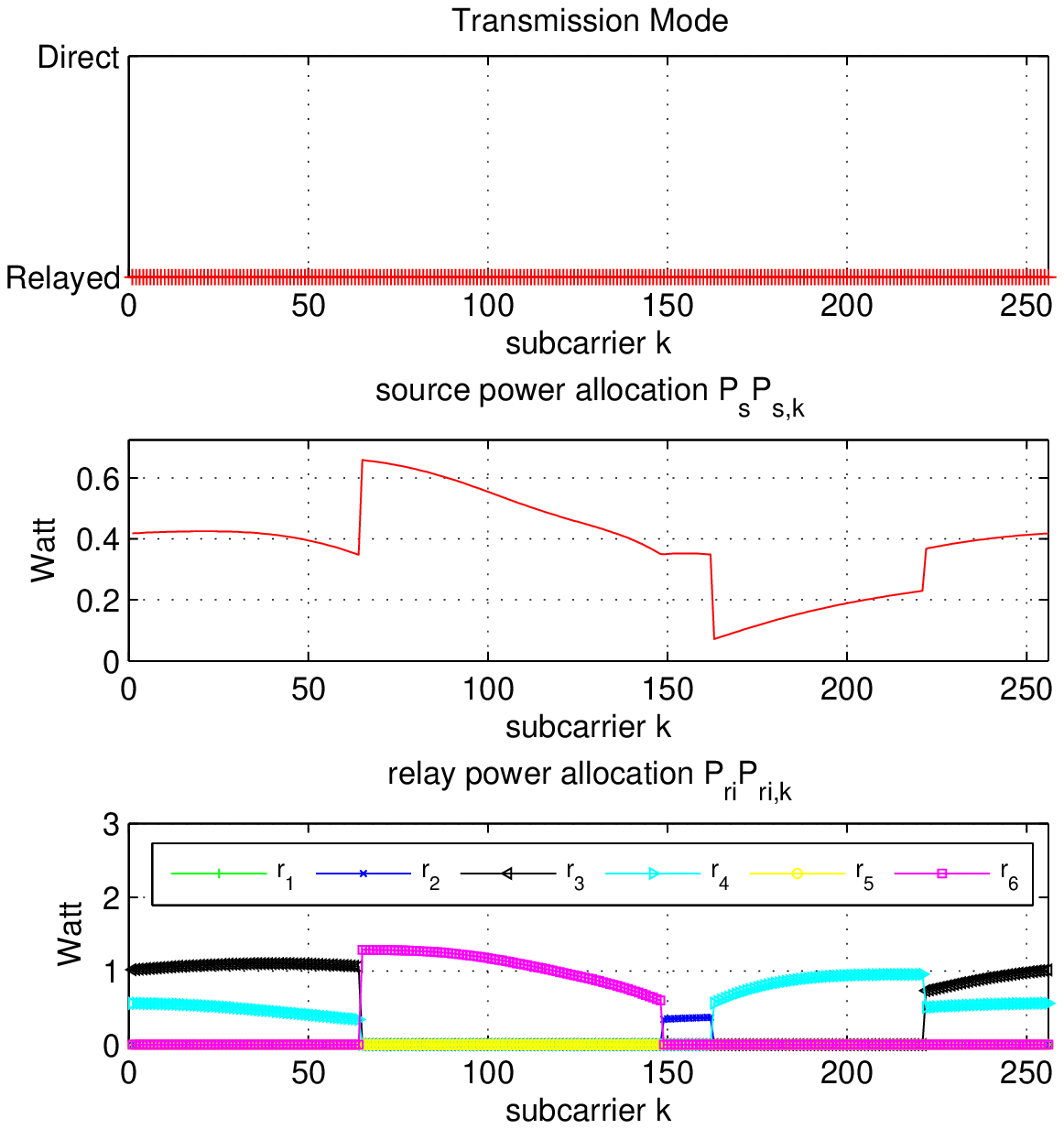}
  \caption{The RA evaluated by {\bf Algorithm \ref{Alg:duality}}
            {for the random channel realization} when $\Ps=\Pri=20$ dBW, $\forall\;\ri\in\Rset$.}  \label{fig:dual}
\end{figure}

\begin{figure}
  \centering
     \includegraphics[width=3.5in]{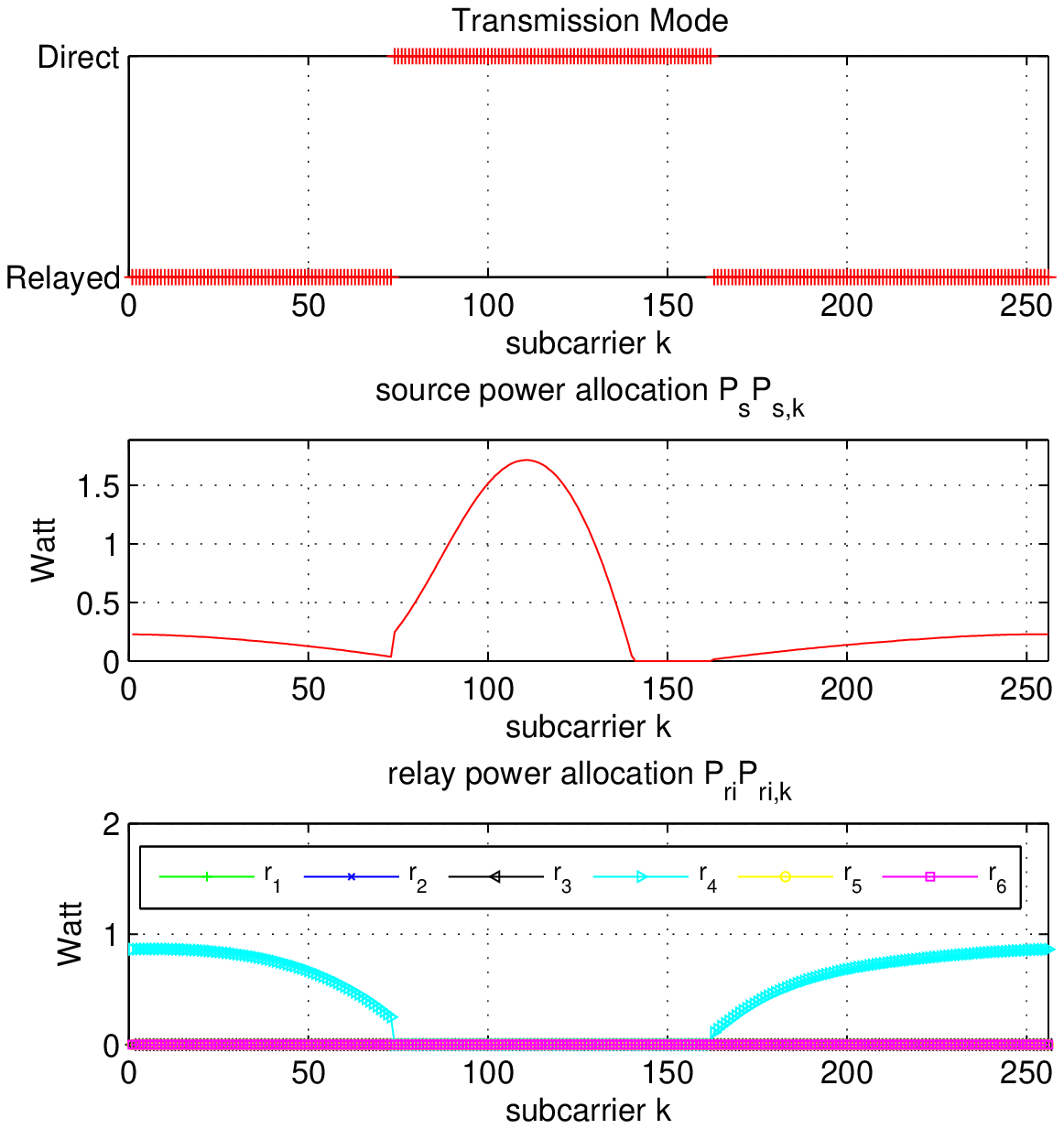}
  \caption{The RA evaluated by {\bf Algorithm \ref{Alg:iterative}}
           {for the random channel realization} when $\Ps=\Pri=20$ dBW, $\forall\;\ri\in\Rset$.}  \label{fig:iterative}
\end{figure}

\begin{figure}
  \centering
     \includegraphics[width=3.5in]{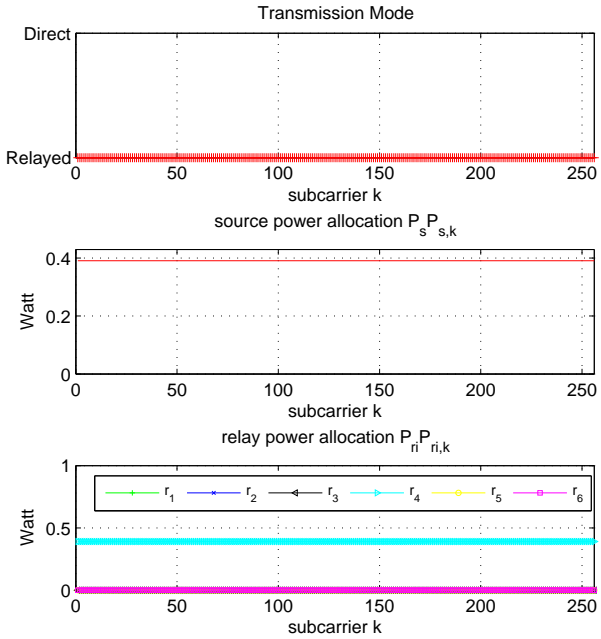}
  \caption{The RA evaluated by the heuristic algorithm {for the random channel realization}
           when $\Ps=\Pri=20$ dBW, $\forall\;\ri\in\Rset$.}  \label{fig:heuristic}
\end{figure}

We have generated a single random channel realization as shown in Figure \ref{fig:oneGain},
and tested the three RA algorithms when $\Ps=\Pri$, $\forall\;\ri\in\Rset$,
and $\Ps$ varies from $10$ to $50$ dBW.
{For the RA evaluated by each algorithm,
the sum rate and the total energy of the source and the relays for transmitting per information bit (TETIB)
are shown in Figure \ref{fig:rate}.
We can see that for each $\Ps$,
either {\bf Algorithm \ref{Alg:duality}} or {\bf Algorithm \ref{Alg:iterative}}
produces a better RA with a higher sum rate and a lower TETIB than the heuristic algorithm,
and {\bf Algorithm \ref{Alg:duality}} produces a better RA than {\bf Algorithm \ref{Alg:iterative}}.
Moreover, the RA evaluated by {\bf Algorithm 1} or {\bf Algorithm 4}
results in a much smaller TETIB than that computed by the heuristic algorithm especially when $\Ps$ is low.}
For the sake of space limitation,
the RAs evaluated by the three algorithms for the random channel realization
only when $\Ps=\Pri=20$ dBW, $\forall\;\ri\in\Rset$,
are shown in Figure \ref{fig:dual}, \ref{fig:iterative}, and \ref{fig:heuristic}, respectively.
When the RA evaluated by {\bf Algorithm 1} is used,
${\rm r_3}$ and ${\rm r_4}$ are enabled to assist relaying simultaneously at a few subcarriers,
while only ${\rm r_4}$ does when the RA computed by {\bf Algorithm 4} or the heuristic algorithm is used.

\begin{figure}
  \centering
     \includegraphics[width=3.5in]{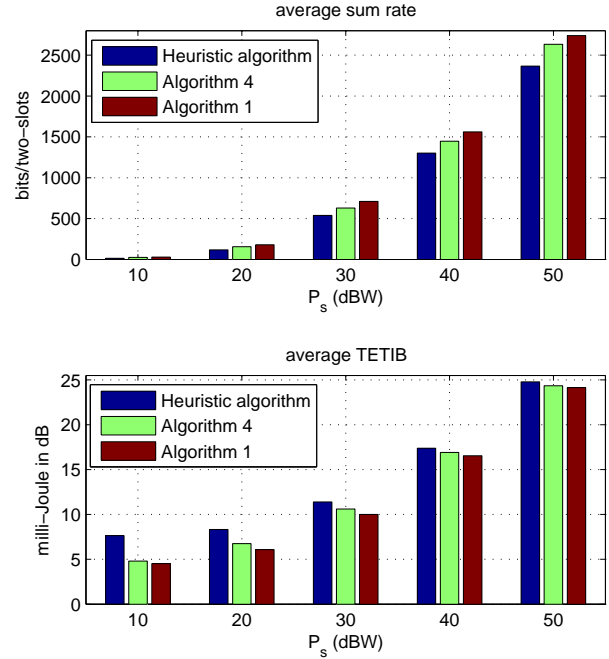}
  \caption{For the RAs evaluated by each algorithm, the average sum rate and the average TETIB
           over 100 random channel realizations
           when $\Ps=\Pri$, $\forall\;\ri\in\Rset$, and $\Ps$ varies from $10$ to $50$ dBW.}  \label{fig:avgrate}
\end{figure}

We have also generated $100$ random channel realizations,
and tested the three RA algorithms when $\Ps=\Pri$, $\forall\;\ri\in\Rset$,
and $\Ps$ varies from $10$ to $50$ dBW.
{For the RAs evaluated by each algorithm, the average sum rate and the average TETIB
are shown in Figure \ref{fig:avgrate}.
It can be seen that for each $\Ps$,
either {\bf Algorithm \ref{Alg:iterative}} or {\bf Algorithm \ref{Alg:duality}} produces better RAs
with a higher average sum rate and a lower average TETIB than the heuristic algorithm,
and {\bf Algorithm \ref{Alg:duality}} produces better RAs than {\bf Algorithm \ref{Alg:iterative}}.
Moreover, the RAs evaluated by {\bf Algorithm 1} or {\bf Algorithm 4}
result in a much smaller average TETIB
than those computed by the heuristic algorithm especially when $\Ps$ is low.}

\section{Conclusion}

We have considered the sum rate maximized RA problem in a point to point OFDM
transmission system assisted by multiple DF relays subject to the individual sum power constraints
of the source and relays.
In particular, one or several relays may cooperate with the source to transmit at every relay-aided subcarrier.
We have proposed two RA algorithms which optimize the assignment of transmission mode and source power
for every subcarrier, as well as the optimum assisting relays and the power allocation to them
for every relay aided subcarrier.
The effectiveness of the two algorithms has been illustrated by numerical experiments.

\appendices

\section{}\label{sec:gap}

When the duality gap is equal to zero,
the duality based algorithm is one of the state-of-the-art methods to
find a globally optimum solution to \eqref{eq:oriprob}.
There exist two important theorems justifying the ability of the duality based algorithm
to find a globally optimum solution to \eqref{eq:oriprob}.
The first (Proposition $3.3.4$ in \cite{Nonlinear-opt}) is that
$\xvec_{\muvec}$ is a globally optimum solution to \eqref{eq:oriprob},
if it is feasible to \eqref{eq:oriprob} and $\muvec(\onevec-g(\xvec_{\muvec}))^T=0$. 
The second (Proposition $5.1.4$ in \cite{Nonlinear-opt}) is that
the $\muvec$ satisfying the aforementioned conditions, denoted by $\optmuvec$,
must minimize $d(\muvec)$ when $\muvec\succcurlyeq\zerovec$.

\begin{figure}[!t]
  \centering
  \subfigure[When $f(\xvec_\pvec)$ is a concave function of $\pvec$]{
     \includegraphics[width=3in]{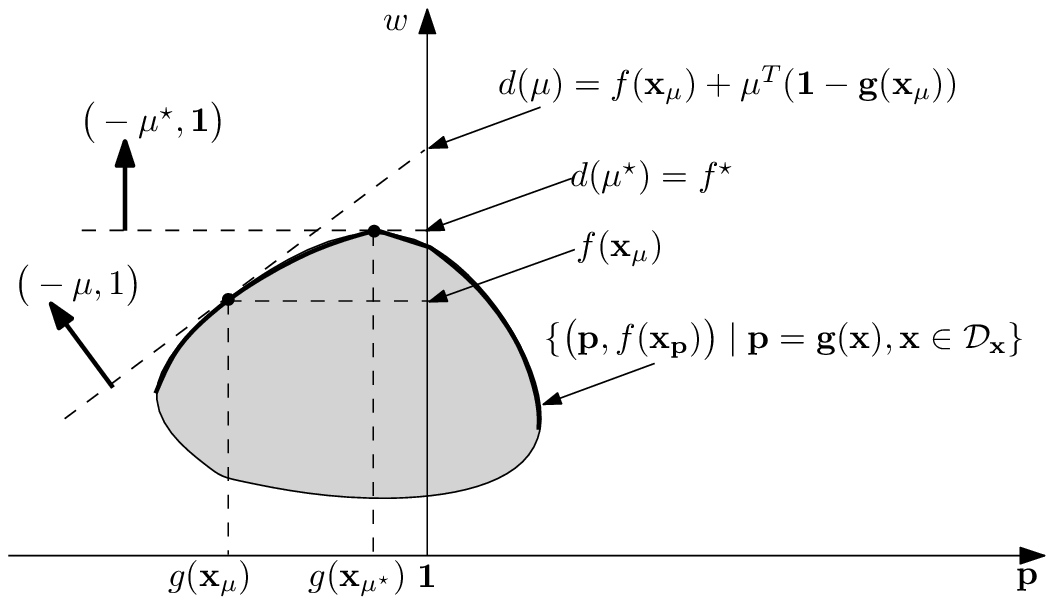}}
  \hspace{0.1in}
  \subfigure[When $f(\xvec_\pvec)$ is not a concave function of $\pvec$]{
     \includegraphics[width=3in]{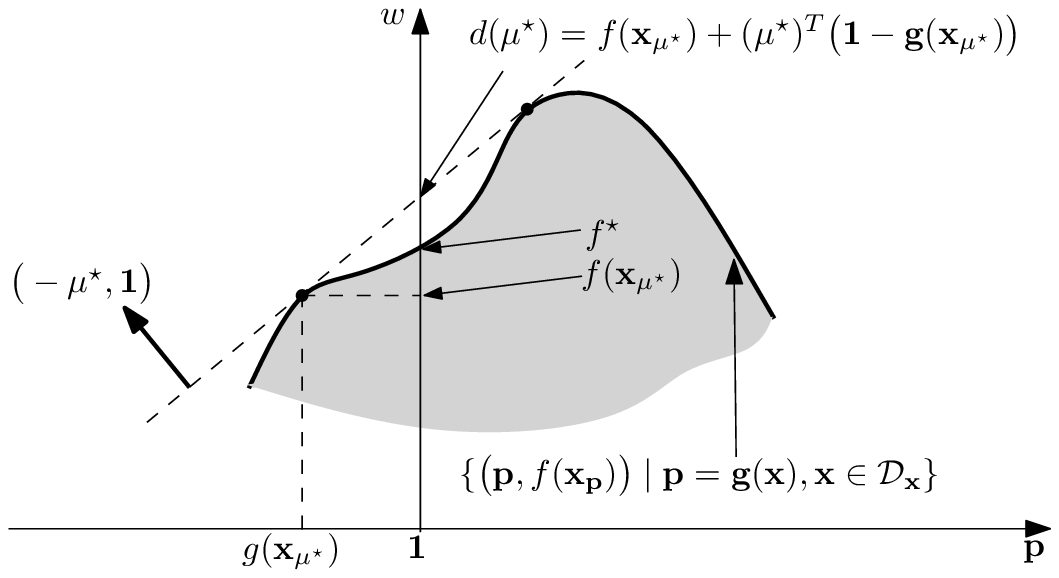}}
  \caption{The visualization of the duality gap in the hyperplane of $[\pvec,w]$. }
  \label{fig:dualitygap}
\end{figure}

In general, the duality gap of \eqref{eq:oriprob} can be studied with a visualization based method
proposed in \cite{Nonlinear-opt}.
Specifically, we can plot a cloud of points collected in the set
$\mathcal{S}=\{[\pvec,w]|\pvec=g(\xvec), w=f(\xvec), \xvec\in\xDom\}$
in the hyperplane of $[\pvec,w]$ shown in Figure \ref{fig:dualitygap}.
Most interestingly, $d(\muvec)$ is equal to the $w$-coordinate of the highest intersection
between the line $\pvec=\onevec$ and a line passing through $\mathcal{S}$
and perpendicular to the vector $[-\muvec,1]$.
As shown in Fig. \ref{fig:dualitygap}, the duality gap is equal to zero
if the $w$-coordinate of the upper border of $\mathcal{S}$ is a concave function of $\pvec$.
Mathematically, a point on the upper border of $\mathcal{S}$ has the coordinate
$(\pvec, f(\xvecp))$, where
\begin{align}
\xvecp  = \arg\max_{\xvec:\xvec\in\xDom,\; g(\xvec) = \pvec} f(\xvec).  \label{eq:borderpoint}
\end{align}

{Based on the same idea first proposed in \cite{Yu06}},
we can show $f(\xvecp)$ is a concave function of $\pvec$ when $K$ is very large.
This is equivalent to show that for any $\theta\in[0,1]$,
\begin{align}
f(\xvecmix) \geq {\theta}f(\xvecpone)+(1-\theta)f(\xvecptwo)  \label{eq:zerogapcondition}
\end{align}
holds for any $\pvecone$ and $\pvectwo$ stacking the sum power of the source and relays.
Note that the above condition can be interpreted in a very interesting way as follows.
Let's adopt the RAs $\xvecpone$ and $\xvecptwo$
in the $\theta$ and $1-\theta$ portions of the whole transmission duration, respectively,
which is called $\theta$ time sharing of $\xvecpone$ and $\xvecptwo$ hereafter.
In this way, an average sum rate ${\theta}f(\xvecpone)+(1-\theta)f(\xvecptwo)$ can be achieved
with an average sum power $\theta\pvecone+(1-\theta)\pvectwo$.
{\it
This means that showing the validity of \eqref{eq:zerogapcondition} for any $\theta\in[0,1]$,
is equivalent to show that the optimum RA for the sum power $\theta\pvecone+(1-\theta)\pvectwo$
provides a higher sum rate than $\theta$ time sharing of $\xvecpone$ and $\xvecptwo$.
}

To this end, we show in the following that when $K$ is very large,
a RA $\xmix\in\xDom$, which is of the sum power $\theta\pvecone+(1-\theta)\pvectwo$ and yields a sum rate
equal to $\theta$ time sharing of $\xvecpone$ and $\xvecptwo$,
can be constructed {\it by $\theta$ spectrum sharing of $\xvecpone$ and $\xvecptwo$,
i.e., taking the entries of $\xvecpone$ at $\theta$ portion of all subcarriers,
and the entries of $\xvecptwo$ at all the other subcarriers to construct $\xmix$.}
More specifically,
we first divide all subcarriers into $S$ subbands,
each consisting of $\NS$ adjacent subcarriers experiencing the same channel conditions.
Suppose $K$ is sufficiently large so that $\NS$ is also very large,
and $\forall\;\theta\in[0,1]$ there exists an integer $N_\theta$ in $\{1,\cdots, \NS\}$
satisfying $\theta\approx\frac{N_\theta}{\NS}$.
Since the subcarriers in each subband experience the same channel conditions,
the entries of $\xvecpone$ (or $\xvecptwo$) at the subcarriers in the same subband should be the same.
Let's construct $\xmix$,
which in every subband adopts for the first $N_\theta$ subcarriers the same entries as in $\xvecpone$,
and for the remaining $\NS-N_\theta$ subcarriers the same entries as in $\xvecptwo$.
We can easily show that the conditions $\xmix\in\xDom$ and
\begin{align}
 g(\xmix) &\approx \theta\pvecone + (1-\theta)\pvectwo,           \nonumber \\
 f(\xmix) &\approx {\theta}f(\pvecone) + (1-\theta)f(\pvectwo)    \nonumber
\end{align}
are all satisfied.
Thus $f(\xvecmix) {\geq} f(\xmix) \approx {\theta}f(\pvecone) + (1-\theta)f(\pvectwo)$ holds.
Therefore, the duality gap of \eqref{eq:oriprob} is equal to zero when $K$ is very large.





\section*{Acknowledgement}

The authors would like to thank the anonymous reviewers
for their valuable comments and suggestions.

\bibliographystyle{IEEEtran}
\bibliography{Relay-ind-bib}

\begin{IEEEbiography}[{\includegraphics[width=1in,height=1.5in,clip,keepaspectratio]{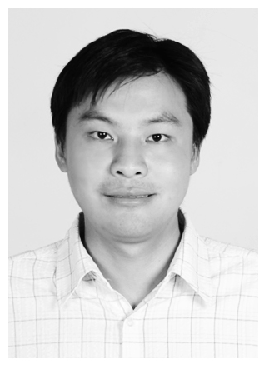}}]{Tao Wang}
received respectively B.E. ({\it summa cum laude}) and DoE degree in electronic engineering
from Zhejiang University, China, in 2001 and 2006,
as well as civil electrical engineering degree ({\it summa cum laude})
from Universit{\'e} Catholique de Louvain (UCL), Belgium, in 2008.
He had multiple research appointments
in Motorola Electronics Ltd. Suzhou Branch, China, since 2000 to 2001,
the Institute for Infocomm Research (I$^2$R), Singapore, since 2004 to 2005,
and Delft University of Technology and Holst Center in the Netherlands since 2008 to 2009.
He is with UCL since 2009.
He has been an associate editor-in-chief for {\it Signal Processing: An International Journal (SPIJ)} since Oct. 2010.
His current research interests are in the optimization of wireless localization systems
with energy awareness, as well as resource allocation algorithms in wireless communication systems.
\end{IEEEbiography}

\begin{IEEEbiography}[{\includegraphics[width=1in,height=1.5in,clip,keepaspectratio]{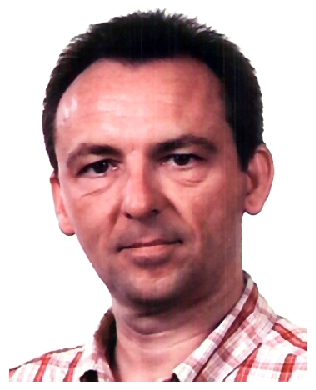}}]{Luc Vandendorpe}
was born in Mouscron, Belgium in 1962. He received the Electrical Engineering degree ({\it summa cum laude})
and the Ph. D. degree from the Université Catholique de Louvain (UCL) Louvain-la-Neuve, Belgium
in 1985 and 1991 respectively. Since 1985, he is with the Communications and Remote
Sensing Laboratory of UCL where he first worked in the field of bit rate reduction techniques
for video coding. In 1992, he was a Visiting Scientist and Research Fellow at the Telecommunications
and Traffic Control Systems Group of the Delft Technical University, The Netherlands, where he worked
on Spread Spectrum Techniques for Personal Communications Systems.
From October 1992 to August 1997, L. Vandendorpe was Senior Research Associate of the Belgian
NSF at UCL, and invited assistant professor. Presently he is Professor and head of
the Institute for Information and Communication Technologies, Electronics and Applied Mathematics.

His current interest is in digital communication systems and more precisely resource
allocation for OFDM(A) based multicell systems, MIMO and distributed MIMO,
sensor networks, turbo-based communications systems, physical layer security and UWB based positioning.
In 1990, he was co-recipient of the Biennal Alcatel-Bell Award from the Belgian NSF
for a contribution in the field of image coding. In 2000 he was co-recipient
(with J. Louveaux and F. Deryck) of the Biennal Siemens Award from the Belgian NSF
for a contribution about filter bank based multicarrier transmission.
In 2004 he was co-winner (with J. Czyz) of the Face Authentication Competition,
FAC 2004. L. Vandendorpe is or has been TPC member for numerous IEEE conferences
(VTC Fall, Globecom Communications Theory Symposium, SPAWC, ICC) and for the Turbo Symposium.
He was co-technical chair (with P. Duhamel) for IEEE ICASSP 2006.
He was an editor of the IEEE Trans. on Communications for Synchronisation and
Equalization between 2000 and 2002, associate editor of the IEEE Trans. on
Wireless Communications between 2003 and 2005, and associate editor of the
IEEE Trans. on Signal Processing between 2004 and 2006. He was chair of the
IEEE Benelux joint chapter on Communications and Vehicular Technology
 between 1999 and 2003. He was an elected member of the Signal Processing for
Communications committee between 2000 and 2005, and an elected member of the Sensor
Array and Multichannel Signal Processing committee of the Signal Processing Society between 2006 and 2008.
Currently, he is an elected member of the Signal Processing for Communications committee.
He is the Editor in Chief for the Eurasip Journal on Wireless Communications and Networking.
He is a Fellow of the IEEE.
\end{IEEEbiography}


\end{document}